\documentclass[final,5p,times,twocolumn,nopreprintline]{elsarticle}

\usepackage{graphicx}
\usepackage{dcolumn}
\usepackage{amsmath,amssymb}
\usepackage{bm}
\usepackage{braket}
\usepackage{appendix}
\usepackage{longtable}
\usepackage{booktabs} 
\usepackage{caption} 
\usepackage[version=4]{mhchem}
\usepackage{footnote}
\usepackage[hidelinks]{hyperref}

\begin{document}

\begin{frontmatter}

\title{Pseudo-hybrid density functional ACBN0 for Hubbard $U$ correction in a numeric atom-centered orbital basis}

\author[skoltech]{Svetlana A. Artiukova\corref{cor1}}
\ead{svetlana.artiukova@skoltech.ru}

\author[skoltech]{Ilia M. Odud}

\author[skoltech]{Sergey V. Levchenko}

\address[skoltech]{Center for Energy Science and Technology, Skolkovo Institute of Science and Technology, 121205 Moscow, Russia}
\cortext[cor1]{Corresponding author}

\begin{abstract}
We present a formulation and implementation of the Agapito--Curtarolo--Buongiorno Nardelli (ACBN0) pseudo-hybrid density functional in a numeric atom-centered orbital basis. The method is realized in the all-electron, full-potential electronic-structure package FHI-aims. The implementation uses a L\"owdin-orthogonalized projector, which improved the stability of the self-consistent ACBN0 iterations for the tested systems. For a benchmark set of materials including metal oxides and nitrides, ACBN0 with the fully localized limit (FLL) as the double-counting treatment reduces the aggregate band gap errors relative to the Perdew--Burke--Ernzerhof (PBE) and strongly constrained and appropriately normed (SCAN) functionals, reaching an accuracy comparable to the Heyd--Scuseria--Ernzerhof functional (HSE06). The computational cost of a single ACBN0 self-consistent-field iteration is comparable to that of the parent functional and substantially lower than that of HSE06. The applicability of the localized-basis implementation to low-dimensional systems is demonstrated by calculations of adsorption and the oxygen evolution reaction (OER) overpotential on a $\beta$-NiOOH(001) surface. On structures relaxed with regularized SCAN (rSCAN), single-point ACBN0 calculations based on rSCAN (ACBN0@rSCAN) with the ``tight'' numerical settings reduce the mean absolute deviation of the reaction-step energies from HSE06 and shift the estimated overpotential toward the HSE06 and Perdew--Burke--Ernzerhof hybrid (PBE0) reference results.
\end{abstract}


\end{frontmatter}


\section{Introduction}

Machine-learning interatomic potentials are increasingly used in materials science, and their reliability depends critically on the quality, size, and diversity of the training data. Such data are commonly generated by density functional theory (DFT), one of the most widely used electronic-structure methods in computational materials science and high-throughput materials discovery. Local and semi-local exchange-correlation (xc) approximations, including the local-density approximation (LDA), generalized-gradient approximation (GGA), and meta-GGA functionals, are frequently used because they offer a favorable balance between computational cost and throughput \cite{behler2016perspective,mueller2016machine}. For strongly correlated materials, however, self-interaction error (SIE) can cause excessive delocalization of states, leading to an inadequate description of charge localization, magnetism, and bonding. Hybrid functionals, such as the Heyd--Scuseria--Ernzerhof functional (HSE06) \cite{heyd2003hybrid,paier2006screened}, partly reduce this error through the inclusion of non-local exact exchange and often provide a more reliable description of localized states. Their high computational cost limits systematic use in materials discovery, especially for large periodic systems and high-throughput or active-learning workflows.

The DFT+$U$ method \cite{anisimov1991band,dudarev1998electron} augments DFT with an on-site Hubbard correction for a selected localized subspace and provides a computationally affordable treatment to improve the description of electronic structure. For bulk materials, DFT+$U$ and its extensions, including DFT+$U$+$J$, have been widely used to describe band gaps \cite{may2020acbn0,lambert2023bandgaps}, formation and defect-formation energies \cite{aykol2014thermochemistry,lambert2024defects}, magnetic states and exchange interactions \cite{keshavarz2018magnetism}, and redox energetics \cite{gautam2018scanU}. Beyond bulk materials, DFT+$U$ has been used to investigate how electron localization affects the electronic, magnetic, and optical properties of low-dimensional systems and surfaces. In Ag-doped GaN nanotubes, DFT-LSDA+$U$ calculations predict a decrease in the band gap with increasing dopant concentration and a stable ferromagnetic state, suggesting potential spintronic applications \cite{naziyev2024gan}. For armchair carbon nanotubes, a systematic scan of $U$ showed that antiferromagnetic order and a gap appear only above a diameter-dependent critical interaction, whereas displacement of the Fermi level by a gate voltage suppresses both quantities \cite{kostyrko2024cnt}. Time-dependent DFT+$U$ (TDDFT+$U$) has also reproduced the evolution from discrete optical excitations in small Ag clusters to plasmonic resonances in nanoparticles containing hundreds of atoms. In that case, a parameter obtained for bulk Ag remained transferable across the investigated size range \cite{chaudhary2024silver}. For oxide surfaces, linear-response Hubbard parameters have been used to investigate oxygen evolution trends \cite{xu2015linearresponse}, while surface-specific $U$ values fitted to X-ray photoelectron shifts have been used to identify adsorbed species \cite{jiang2023surfaceU}. In bilayer chromium trihalides, the predicted interlayer magnetic coupling depends on both $U$ and the double-counting treatment \cite{sarkar2021magnetic}. Large-scale calculations for two-dimensional materials further show that the effects of $U$ on band gaps, magnetic moments, exchange coupling, and magnetic anisotropy depend strongly on the material \cite{pakdel2025twod}. These studies demonstrate the relevance of Hubbard $U$ correction to low-dimensional electronics, spintronics, catalysis, and nanoplasmonics, but they also expose a common methodological problem in DFT+$U$. The Hubbard parameter is often transferred from a bulk or monolayer reference, fitted to an experimental observable, or scanned as an external variable. Although such transfer can be successful, as reported for Ag clusters, it cannot be assumed when reduced coordination, curvature, defects, adsorption, or dielectric screening changes the localization of the corrected states. Chemically inequivalent atoms of the same species can have different Hubbard parameters. A first-principles, atom-resolved evaluation of the Hubbard parameters is therefore important for predictive calculations.

Several first-principles schemes have been proposed to address this problem. In the linear-response method \cite{cococcioni2005linear}, localized orbitals are perturbed by small shifts of their on-site potential, and the changes in their occupations yield $U=\chi_0^{-1}-\chi^{-1}$, where $\chi_0$ and $\chi$ are the non-interacting and interacting response functions of the correlated subspace. In the constrained random-phase approximation (cRPA) \cite{aryasetiawan2006calculations}, a partially screened Coulomb interaction is calculated for the target low-energy subspace while screening processes internal to that subspace are excluded. Although physically well founded, these methods can be computationally demanding.

Agapito--Curtarolo--Buongiorno Nardelli (ACBN0) \cite{agapito2015reformulation} provides a less expensive alternative by evaluating the Hubbard parameters self-consistently from electron-repulsion integrals and renormalized density matrices in the chosen localized subspace. It was originally implemented as an outer self-consistency loop around separate DFT+$U$ calculations performed with the
\textit{Quantum ESPRESSO} software package using a plane-wave basis and
norm-conserving pseudopotentials
\cite{agapito2015reformulation,giannozzi2009quantum}. The Hubbard
parameters obtained from a given electronic density were passed to the
subsequent DFT+$U$ calculation, and this sequence was repeated until the
parameters converged. Tancogne-Dejean, Oliveira, and Rubio subsequently implemented ACBN0 in the \textit{Octopus} pseudopotential real-space software package \cite{tancogne2017selfconsistent,tancogne2020octopus}. In that implementation, the Hubbard parameters are evaluated during the self-consistent-field (SCF) energy minimization, without an external sequence of separate DFT+$U$
calculations. The Kohn--Sham states are represented on a real-space grid, while the Hubbard subspaces are constructed from localized pseudo-atomic orbitals supplied with the norm-conserving pseudopotentials and represented on atom-centered subgrids \cite{tancogne2017selfconsistent}.

Several applications of ACBN0 have been reported in the literature. Representative applications to bulk materials include the
study by Gopal \textit{et al.}, who reported improved electronic,
vibrational, and structural properties of Zn- and Cd-based chalcogenides relative to the Perdew--Burke--Ernzerhof (PBE) functional \cite{perdew1996generalized,gopal2015acbn0}. May and Kolpak
subsequently applied self-consistent ACBN0 to first-row transition metal perovskites and obtained more accurate crystal structures and band gaps than with PBE or conventional DFT+$U$ using
Hubbard parameters taken from the literature \cite{may2020acbn0}.
Self-consistent ACBN0 has also been combined with magnetic and thermoelectric measurements in a study of the Co spin states and Co/Nb arrangements in Ba$_2$CoNbO$_6$ \cite{ba2conbo6}. Further applications to Ruddlesden--Popper oxides examined how the La/Sr distribution affects
the electronic and magnetic properties of LaSrFeO$_4$ \cite{fazlizhanova2024lasrfeo4}, and how the cation arrangement changes
the electronic structure, magnetic order, and site-resolved Hubbard parameters in mixed Co--Fe systems \cite{fazlizhanova2024config,fazlizhanova2025lasrcofe}. Applications to adsorption have also been reported. Huang \textit{et al.} studied hydroxyl adsorption on the polar surfaces of monolayer $\alpha$-In$_2$Se$_3$ using first-principles Hubbard parameters \cite{huang2020ferroelectrics}. Applying ACBN0 to the substrate improved the adsorption energies relative to PBE, although the results remained qualitatively different from HSE06. When the self-consistent Hubbard correction was also applied to the O-$2p$ states of the open-shell hydroxyl adsorbate, the calculated adsorption energies became close to the HSE06 results. This agreement highlights
the importance of treating the localized states of both the substrate and the molecular adsorbate consistently.

The aim of this work is to formulate and assess ACBN0 in a basis of numeric atom-centered orbitals. Unlike a plane-wave expansion, an atom-centered basis requires no additional orbital basis functions when vacuum regions are enlarged, which is advantageous for surfaces, clusters, and molecules. Compared with previous pseudopotential-based implementations, the present formulation uses the all-electron, full-potential FHI-aims framework to explicitly describe both core and valence states without pseudopotentials \cite{blum2009ab}. The construction of the localized subspace for the Hubbard $U$ correction in a non-orthogonal atom-centered orbital basis nevertheless remains a distinct methodological problem, since the projector and basis choices can affect both numerical stability and predictive accuracy. We address this problem using a L\"{o}wdin-orthogonalized projector and investigate the following related questions: whether it stabilizes the ACBN0 iterations across a chemically diverse test set, how the results depend on the size of the numeric atom-centered orbital basis, how the fully localized limit (FLL), around-mean-field (AMF), and Petukhov double-counting treatments affect the predictions \cite{petukhov2003correlated}, and how accurately the method describes bulk band gaps and magnetic moments relative to experimental and hybrid functional reference data. Adsorption and OER calculations on $\beta$-NiOOH(001) extend this assessment to chemically heterogeneous surfaces, testing the consistent application of atom-resolved corrections to the substrate, adsorbates, and molecular references against hybrid functional calculations.

\section{Hubbard $U$ correction in a numeric atom-centered orbital basis}\label{sec:projectors}
Before introducing the self-consistent evaluation of the Hubbard parameters via the ACBN0 functional, we summarize the DFT+$U$ framework in a numeric atom-centered orbital basis. The physical meaning of a Hubbard correction depends on the definition of the localized subspace and therefore on the projection operators. We then describe the L\"{o}wdin projector used in the present implementation.

The Hubbard correction operator $\hat{V}^{\sigma}_U$ in the DFT+$U$ method is defined by projectors $\hat{P}$ onto a chosen localized subspace:
\begin{equation}
    \hat{H}^{\sigma} = \hat{H}^{\sigma}_{DFT}+\hat{V}^{\sigma}_U,
\end{equation}
\begin{equation}\label{Hamiltonian_correction}
        \hat{V}^{\sigma}_U = \sum_{\{I,n,l\}}\sum_{m,m' \in \{I,n,l\}} V^{\sigma}_{mm'} \hat{P}_{mm'},
  \end{equation}
where $\hat{P}_{mm'}$ is a projector component for the localized subspace defined by the electronic shell with quantum numbers $(n,l)$ for atom $I$. Here and below, whenever indices $m,m',m'',\ldots$ occur, they refer to the same corrected subspace $(I,n,l)$; the shell labels are therefore omitted from matrices, projector components, and potential coefficients. Greek indices $\mu,\nu,\mu',\nu',\ldots$ run over the complete basis.

The coefficients $V^{\sigma}_{mm'}$ depend on the double-counting treatment. The AMF limit \cite{czyzyk1994local} is formulated relative to the average shell occupation, whereas FLL \cite{czyzyk1994local,PhysRevB.48.16929} is appropriate for localized states with occupations close to integers. Petukhov {\em et al.} \cite{petukhov2003correlated} proposed a linear interpolation between AMF and FLL, controlled by a mixing parameter $\alpha$ determined self-consistently from the orbital occupation matrix. They concluded that this approach is most appropriate for moderately correlated metals, where the orbital occupancies are neither uniform nor fully localized, while for weakly correlated metals the method remains fundamentally inadequate because it still does not describe dynamical fluctuations. In the general case,
\begin{equation}\label{V}
\begin{aligned}
    V^{\sigma}_{mm'} = -U^{I,n,l}\Bigg\{n_{mm'}^{\sigma}
    &-\left[(1-\alpha^{I,n,l})N^{\sigma}\right.\\[-0.3ex]
    &\left.\quad+\frac{\alpha^{I,n,l}}{2}\right]\delta_{mm'}\Bigg\},
\end{aligned}
\end{equation}
where $U^{I,n,l}=\bar U^{I,n,l}-\bar J^{I,n,l}$ is the Hubbard parameter in the Dudarev formulation \cite{dudarev1998electron}. For periodic calculations,
\begin{equation}\label{eq:lowd_occ_matrix}
    n^{\sigma}_{mm'} = \sum_{i,\bm{k}}w_{\bm{k}}f_{i\bm{k}}\bra{\psi^{\sigma}_{i,\bm{k}}}\hat{P}^{\bm{k}}_{mm'}\ket{\psi^{\sigma}_{i,\bm{k}}},
\end{equation}
is the occupation matrix for the corresponding localized subspace. Here, $\psi^{\sigma}_{i,\bm{k}}$ is a Kohn--Sham state, $w_{\bm{k}}$ is the $\bm{k}$-point weight, and $f_{i\bm{k}}$ is the corresponding occupation number.

According to Ref. \cite{petukhov2003correlated}, the mixing parameter $\alpha$ is defined for each $(I,n,l)$ shell, combining both spin channels, as follows:
\begin{equation}\label{mixing2}
    \alpha^{I,n,l} = \frac{\sum_{\sigma} \sum_{m,m'}\delta n^{\sigma}_{mm'} \delta n^{\sigma}_{m'm}}{(2l+1)\sum_{\sigma} N^{\sigma}(1-N^{\sigma})},
\end{equation}
where $N^{\sigma}=\frac{1}{2l+1}\sum_{m}n^{\sigma}_{mm}$ is the average occupation in the $(I,n,l,\sigma)$ shell, and
\begin{equation}\label{mixing3}
    \delta n^{\sigma}_{mm'}  =  n^{\sigma}_{mm'} - N^{\sigma} \delta_{mm'}.
\end{equation}

The definition of the localized Hubbard subspace is one of the central choices in DFT+$U$. Let $Q$ denote a selected subspace and let $\{\ket{\tilde{\varphi}_m}\}$ be an orthonormal basis for this subspace. Its projector and projected population are
\begin{equation}\label{projected_population}
    \hat P_Q=\sum_{m\in Q}\ket{\tilde{\varphi}_m}\bra{\tilde{\varphi}_m},
    \qquad
    N_Q=\sum_{\sigma}\operatorname{Tr}\!\left[\hat\rho^{\sigma}\hat P_Q\right],
\end{equation}
where $\hat\rho^{\sigma}$ is the one-particle density operator for spin channel $\sigma$. A population analysis satisfies the electron-number completeness condition if the sum of the populations obtained by applying the construction to all basis functions equals the total number of electrons. In FHI-aims, the Kohn--Sham states are expanded in numeric atom-centered orbitals located on different atoms throughout the structure. These functions generally overlap in space and are therefore not mutually orthogonal, so the assignment of the electron density to individual basis functions is not unique. Kick \textit{et al.}~\cite{FHIaimsDFT+U} described the on-site and symmetrized Mulliken projector definitions used for DFT+$U$ in FHI-aims. If extended to all basis functions, the on-site construction does not in general preserve the total electron number in a non-orthogonal basis, whereas the symmetrized Mulliken construction and the L\"{o}wdin construction used here do. The symmetrized Mulliken construction uses the dual basis functions $\{\ket{\tilde{\varphi}_m}\}$ and is defined by Eqs.~(\ref{mulliken1}) and (\ref{mulliken2}):

\begin{equation}\label{mulliken1}
    \ket{\tilde{\varphi}_{m}} = \sum_{\mu} \left[S^{-1}\right]_{\mu m} \ket{\varphi_{\mu}},
\end{equation}
\begin{equation}\label{mulliken2}
    \hat{P}_{mm'} = \frac{1}{2}
\left( \ket{\tilde{\varphi}_{m'}} \bra{\varphi_m}+\ket{\varphi_{m'}}\bra{\tilde{\varphi}_m}\right).
\end{equation}

This construction preserves the total electron number, but its components do not form a set of mutually orthogonal projectors. L\"{o}wdin orthogonalization, discussed in Sec.~\ref{Lowdin_subsec}, instead produces an orthonormal set and therefore defines a conventional orthogonal projection onto the selected subspace.

However, defining the basis for this localized subspace is not a trivial task. The numeric atom-centered orbital basis consists of a \textit{minimal} basis, i.e., numerical orbitals of spherically symmetric free atoms, supplemented by additional radial functions with the same angular form. These additional functions improve the description of orbital polarization, chemical bonding, hybridization, and spatially extended components of the electronic states. In the calculations presented here, the localized Hubbard subspace for a selected $(n,l)$ shell is defined from the \textit{minimal}, i.e., free-atomic basis. Higher-tier functions remain part of the Kohn--Sham basis, but are not selected as additional independent Hubbard orbitals. Consequently, $N_Q$ contains only the population of the selected subspace and excludes other basis functions with the same $(n,l)$ character. With this minimal-basis definition, the Mulliken construction retains the selected functions in their original non-orthogonal form and symmetrically partitions their overlap with higher-tier functions and functions on neighboring atoms. The L\"{o}wdin construction instead orthogonalizes the complete basis before selecting the functions associated with the minimal shell. The selected L\"{o}wdin functions are therefore orthogonal to the remaining basis functions, and their occupations do not require an explicit partition of shared overlap. Higher-tier functions can nevertheless contribute to the selected L\"{o}wdin orbitals through the full-basis orthogonalization. This choice of subspace contributes to the observed dependence of the results on the basis. Alternative definitions based on linear combinations of several radial functions, maximally localized Wannier functions, or natural atomic orbitals may reduce this basis dependence.

\subsection{L\"{o}wdin projection functions in periodic calculations}\label{Lowdin_subsec}

L\"{o}wdin orthogonalization is applied to the complete finite basis at each $\bm{k}$ point; the functions defining the Hubbard subspace are then selected from this orthonormal representation. In FHI-aims, the Bloch-like basis functions $\{\varphi_\mu^{\bm{k}}\}$ are expressed via atomic orbitals $\{\varphi_\mu\}$ as follows:
\begin{equation}\label{basis_k}
    \varphi_{\mu}^{\bm{k}}(\bm{r}) = \sum_{\bm{R}} e^{i\bm{k}\cdot\bm{R}} \varphi_\mu(\bm{r}-\bm{R}),
\end{equation}
where the sum formally runs over all lattice vectors $\bm{R}$. The orthogonalized basis functions are expressed as follows:
\begin{equation}\label{Lowdin_orth}
    \tilde{\varphi}_m^{\bm{k}} = \sum_{\mu} \varphi_{\mu}^{\bm{k}}\left[\left(S^{\bm{k}}\right)^{-1/2}\right]_{\mu m}.
\end{equation}
Here $S^{\bm{k}}_{\mu\nu} = \braket{\varphi_{\mu}^{\bm{k}}|{\varphi_{\nu}^{\bm{k}}}}$ is the overlap matrix at the $\bm{k}$ point.
The occupation matrix $n_{mm'}^{\sigma}$ for the spin channel $\sigma$ is expressed in Eq. (\ref{eq:lowd_occ_matrix}), with
\begin{equation}  
\hat{P}^{\bm{k}}_{mm'} = \ket{\tilde{\varphi}_{m'}^{\bm{k}}}\bra{\tilde{\varphi}_{m}^{\bm{k}}}.
\label{Lowdin_proj}
\end{equation}
The \textit{i}-th Kohn--Sham state is expressed in terms of the atomic orbitals and Bloch-like basis functions as follows:
\begin{equation}\label{KS-state}
\psi_{i,\bm{k}}^{\sigma} (\bm{r})= \sum_{\mu,\bm{R}} c_{\mu i}^{\bm{k} \sigma} e^{i\bm{k}\cdot\bm{R}} \varphi_{\mu}(\bm{r}-\bm{R}) = \sum_{\mu} c_{\mu i}^{\bm{k} \sigma} \varphi_{\mu}^{\bm{k}}(\bm{r}).
\end{equation}
The projections $\braket{\psi_{i,\bm{k}}^{\sigma}|\varphi_{m}^{\bm{k}}}$ and their complex-conjugate counterparts appearing in Eq.~(\ref{eq:lowd_occ_matrix}) formally contain two infinite sums over the lattice vectors. These sums can be rearranged so that one becomes a sum of equivalent contributions from the periodically repeated unit cells. Retaining the contribution per unit cell gives the occupation matrix
\begin{equation}
\begin{aligned}
    n_{mm'}^{\sigma}={}&\sum_{i,\bm{k}} w_{\bm{k}} f_{i \bm{k}}
    \sum_{\mu,\nu} c^{\bm{k} \sigma *}_{\mu i}
    \left[\left(S^{\bm{k}}\right)^{1/2}\right]_{\mu m'}\\[-0.3ex]
    &\times\left[\left(S^{\bm{k}}\right)^{1/2}\right]_{m\nu}
    c^{\bm{k} \sigma}_{\nu i}.
\end{aligned}
\end{equation}

\subsection{Hamiltonian and total-energy correction in the L\"{o}wdin representation}
The L\"{o}wdin projector component used below follows directly from Eqs.~(\ref{Lowdin_orth}) and (\ref{Lowdin_proj}).
Using Eq.~(\ref{V}), the Hamiltonian matrix correction can then be expressed as
\begin{equation}
    \Delta H_{\mu\nu}^{\sigma,\bm{k}}=\sum_{m,m'}\left[\left(S^{\bm{k}}\right)^{1/2}\right]_{\mu m'}V_{mm'}^{\sigma}\left[\left(S^{\bm{k}}\right)^{1/2}\right]_{m\nu}.
\end{equation}

The FHI-aims total-energy expression contains a Kohn--Sham eigenvalue sum in addition to the direct density-functional terms [see Eq.~(7) of Ref.~\cite{blum2009ab}]. Because this eigenvalue sum already contains the Hubbard contribution, that contribution must be removed once when adding the Hubbard energy correction $E_U$, leading to
\begin{equation}
    E^{\mathrm{FHI\text{-}aims}}_{U} = E_U - \sum_{i,\bm{k},\sigma}
    w_{\bm{k}} f_{i\bm{k}}\varepsilon_{i\bm{k}}^{U,\sigma},
\end{equation}
where $w_{\bm{k}}$ is the $\bm{k}$-point weight, $f_{i\bm{k}}$ is the occupation number, and $\varepsilon_{i\bm{k}}^{U,\sigma}$ is the expectation value of the operator in Eq.~(\ref{Hamiltonian_correction}):
\begin{equation}
    \varepsilon_{i\bm{k}}^{U,\sigma}=\bra{\psi^{\sigma}_{i,\bm{k}}}\hat{V}_{U}^{\sigma}\ket{\psi^{\sigma}_{i,\bm{k}}}.
\end{equation}

\section{First-principles Hubbard $U$ correction using the ACBN0 pseudo-hybrid density functional}\label{sec:ACBN0inFHIaims}

The present implementation evaluates the ACBN0 Hubbard parameters directly within the single-point calculation in the all-electron, full-potential FHI-aims electronic-structure package. The numeric atom-centered orbital basis provides localized functions from which the Hubbard subspace can be constructed. In conjunction with the L{\"o}wdin projector introduced in Section~\ref{sec:projectors}, this choice was found to improve the numerical stability of the projected occupations and of the coupled SCF--ACBN0 cycle in the calculations performed here.

According to the original algorithm, the Hubbard parameter in ACBN0, $U^{I,n,l}=\bar U^{I,n,l}-\bar J^{I,n,l}$, is determined through electron-repulsion integrals $(mm'|m''m''')$ and renormalized density matrices. The averaged Coulomb and exchange interactions are given by \cite{tancogne2017selfconsistent}:
\begin{equation}\label{eq:acbn0_Ubar}
\bar U^{I,n,l} =
\frac{\sum_{\{m\}}\sum_{\alpha,\beta}
\bar P_{mm'}^{\alpha}\bar P_{m''m'''}^{\beta}(mm'|m''m''')}
{\sum_{m\ne m'}\sum_{\alpha}N_m^{\alpha}N_{m'}^{\alpha}
+\sum_{m,m'}\sum_{\alpha}N_m^{\alpha}N_{m'}^{-\alpha}},
\end{equation}
\begin{equation}\label{eq:acbn0_Jbar}
\bar J^{I,n,l} =
\frac{\sum_{\{m\}}\sum_{\alpha}
\bar P_{mm'}^{\alpha}\bar P_{m''m'''}^{\alpha}(mm'''|m''m')}
{\sum_{m\ne m'}\sum_{\alpha}N_m^{\alpha}N_{m'}^{\alpha}}.
\end{equation}
Here $m,m',m'',m'''$ belong to the same corrected shell $(I,n,l)$, so the shell indices are omitted from the matrices; $\sum_{\{m\}}$ denotes summation over all four indices. The Greek indices $\alpha$ and $\beta$ denote spin channels, and $-\alpha$ denotes the spin channel opposite to $\alpha$. The renormalized density matrix $\bar P_{mm'}^{\alpha}$ and the state-resolved renormalization factor $\bar N_{\psi_{i,\bm{k}}}^{\alpha}$ are defined in Eqs.~(7) and (8) of the same reference, while the orbital occupations satisfy $N_m^{\alpha}=n_{mm}^{\alpha}$. The sum $\sum_{\{I\}}$ in the renormalization factor runs over all atoms of the required chemical species within the unit cell.

The corresponding expressions can be written in terms of the coefficient-space density kernel $\rho^{\sigma,i\bm{k}}_{\mu\nu}=c^{\bm{k}\sigma}_{\mu i}c^{\bm{k}\sigma *}_{\nu i}$:

\begingroup
\small
\noindent\textit{L\"{o}wdin projector:}
\begin{align}\label{Lowdin_acbn0}
n_{mm'}^{\sigma}
&=\sum_{i,\bm{k}}w_{\bm{k}}f_{i\bm{k}}\sum_{\mu,\nu}
\left[\left(S^{\bm{k}}\right)^{1/2}\right]_{m\mu}
\rho^{\sigma,i\bm{k}}_{\mu\nu}
\left[\left(S^{\bm{k}}\right)^{1/2}\right]_{\nu m'},\nonumber\\
\bar N_{\psi_{i,\bm{k}}}^{\sigma}
&=\sum_{\{I\}}\sum_m\sum_{\mu,\nu}
\left[\left(S^{\bm{k}}\right)^{1/2}\right]_{m\mu}
\rho^{\sigma,i\bm{k}}_{\mu\nu}
\left[\left(S^{\bm{k}}\right)^{1/2}\right]_{\nu m},\\
\bar P_{mm'}^{\sigma}
&=\sum_{i,\bm{k}}w_{\bm{k}}f_{i\bm{k}}\bar N_{\psi_{i,\bm{k}}}^{\sigma}
\sum_{\mu,\nu}
\left[\left(S^{\bm{k}}\right)^{1/2}\right]_{m\mu}
\rho^{\sigma,i\bm{k}}_{\mu\nu}
\left[\left(S^{\bm{k}}\right)^{1/2}\right]_{\nu m'}.\nonumber
\end{align}

\noindent\textit{Mulliken projector:}
\begin{align}\label{mulliken_acbn0}
n_{mm'}^{\sigma}
&=\frac{1}{2}\sum_{i,\bm{k}}w_{\bm{k}}f_{i\bm{k}}\sum_{\mu}
\left(\rho^{\sigma,i\bm{k}}_{\mu m'}S^{\bm{k}}_{m\mu}
+\rho^{\sigma,i\bm{k}}_{m\mu}S^{\bm{k}}_{\mu m'}\right),\nonumber\\
\bar N_{\psi_{i,\bm{k}}}^{\sigma}
&=\frac{1}{2}\sum_{\{I\}}\sum_m\sum_{\mu}
\left(\rho^{\sigma,i\bm{k}}_{\mu m}S^{\bm{k}}_{m\mu}
+\rho^{\sigma,i\bm{k}}_{m\mu}S^{\bm{k}}_{\mu m}\right),\\
\bar P_{mm'}^{\sigma}
&=\frac{1}{2}\sum_{i,\bm{k}}w_{\bm{k}}f_{i\bm{k}}\bar N_{\psi_{i,\bm{k}}}^{\sigma}
\sum_{\mu}
\left(\rho^{\sigma,i\bm{k}}_{\mu m'}S^{\bm{k}}_{m\mu}
+\rho^{\sigma,i\bm{k}}_{m\mu}S^{\bm{k}}_{\mu m'}\right).\nonumber
\end{align}
\endgroup

The self-consistent ACBN0 implementation follows these steps:
\begin{enumerate}
    \item The corrected atoms and shells, the type of projector, and the double-counting treatment are read from the input. Initial values of $\bar U$ and $\bar J$ are assigned. When both are zero, the first electronic step uses the parent xc functional.
    \item The on-site electron-repulsion integrals for the selected shell functions are evaluated with the resolution-of-identity localized variant (RI-LVL) \cite{ihrig2015accurate} and stored for reuse during the self-consistent calculation.
    \item The Kohn--Sham equations are solved using the parent xc functional and the current Hubbard correction. The resulting eigenvectors are used to construct the shell-resolved occupation matrices, state renormalization factors, and renormalized density matrices.
    \item The renormalized density matrices are contracted with the stored electron-repulsion integrals to obtain $\bar U^{I,n,l}$ and $\bar J^{I,n,l}$ independently for every corrected atomic shell. For an $s$ shell, $\bar J$ is zero and $U=\bar U$; otherwise $U=\bar U-\bar J$.
    \item The updated $U$ values are passed to the corresponding Hubbard potential and energy correction for the next SCF iteration. The procedure is repeated until the SCF criteria are satisfied. 
\end{enumerate}

The present implementation connects the self-consistent ACBN0 update to the double-counting treatments available in FHI-aims, including the Petukhov mixing scheme described above. This flexibility is relevant, for example, for chemically heterogeneous calculations containing localized and more delocalized components, but the selected double-counting treatment must be stated because it affects the total-energy differences. In the surface calculations below, the clean slab, adsorbate-covered slabs, and isolated molecular references are treated consistently using Petukhov mixing.

The present implementation is self-consistent but not fully variational with respect to the density dependence of the ACBN0 interactions. The parameters $\bar U[\rho]$ and $\bar J[\rho]$ are updated together with the electronic density within the SCF cycle, while the Kohn--Sham potential contains the conventional DFT+$U$ contribution evaluated at their current values. Additional terms associated with the functional derivatives $\delta\bar U/\delta\rho$ and $\delta\bar J/\delta\rho$ are not included. The converged solution is therefore a fixed point of the coupled density--ACBN0 iteration, but it is not obtained by varying a total-energy functional that includes the complete density dependence of $\bar U$ and $\bar J$. Consequently, total-energy differences calculated with the current implementation contain additional approximations, particularly when the compared systems have different chemical environments and different self-consistent Hubbard parameters.

The choice of the projector discussed above is particularly important for \textit{ab initio} Hubbard $U$ calculations with ACBN0, because nonphysical occupations would directly affect the self-consistent $\bar U$ and $\bar J$ parameters and could destabilize the coupled SCF--ACBN0 cycle. A fixed Hubbard parameter can be combined consistently with any projector, provided that an appropriate $U$ value is chosen for it. In the bulk test calculations, the Mulliken projector produced unstable ACBN0 iterations most often for oxygen 2$p$ shells. The numerical results reported below therefore use the L\"{o}wdin projector.

\section{Computational details} 
The computational study consists of two parts: tests for bulk semiconductors and insulators, and an examination of adsorption and OER energetics on the $\beta$-NiOOH(001) surface. The FHI-aims ``light'' and ``tight'' settings denote complete sets of numerical defaults that differ in their integration grids and, most importantly for the present comparison, in the number of numeric atom-centered basis functions \cite{blum2009ab}. The ``tight'' settings include a larger set of functions from higher tiers than the ``light'' settings. 

The bulk benchmark was chosen as a diagnostic set spanning different shell fillings, localization regimes, magnetic states, and mechanisms controlling the band gap. The $3d$ transition metal oxides range from the nominally $d^0$ Ti compounds through antiferromagnetic oxides with partially filled $d$ shells to the $d^9$ and $d^{10}$ Cu compounds. This series spans different degrees of $3d$ localization and metal-oxygen hybridization, allowing examination of how the Hubbard correction affects band gaps across different shell fillings. The alkaline earth and post-transition metal oxides provide complementary closed-shell cases for examining the response of nearly filled Hubbard subspaces and the differences between double-counting treatments in this limit. AlN and GaN extend the benchmark to predominantly $sp$-bonded semiconductors, allowing examination of the effect of ACBN0 when the band edges are not dominated by transition metal $d$ states. The set is not intended as a statistically exhaustive benchmark. It supports conclusions about the response of ACBN0 to these distinct electronic regimes, while the quantitative performance remains material dependent.

For the five magnetic materials under study (Cr$_2$O$_3$, MnO, CoO, NiO, CuO), the antiferromagnetic (AFM) ordering of magnetic moments was used. The atomic structures were relaxed with the GGA functional PBE using the Broyden--Fletcher--Goldfarb--Shanno (BFGS) algorithm and the FHI-aims ``tight'' numerical settings. The full set of materials, lattice parameters, and $\bm{k}$-point meshes is given in Supporting Information (SI) Table~S1. The meta-GGA SCAN functional \cite{sun2015strongly}, the hybrid functional HSE06 \cite{heyd2003hybrid,paier2006screened}, and ACBN0 calculations with PBE and SCAN as underlying xc functionals were performed as single-point calculations on the same PBE-relaxed geometries to isolate differences arising from the electronic-structure description. The corrected orbitals for every material are specified in SI Tables~S6--S9, which also report the self-consistently converged Hubbard $U$ parameters. The calculations were spin-polarized with collinear spins, used Gaussian occupation broadening of 0.01~eV, and evaluated local magnetic moments by Mulliken partitioning. The SCF thresholds were $10^{-3}$~eV for the sum of Hamiltonian eigenvalues, $10^{-5}$~eV for the total energy, and $10^{-4}$ for the charge density. During structural relaxation, forces were evaluated with an SCF accuracy of $10^{-4}$~eV/\AA, and the BFGS optimization was terminated when the maximum force component was below $10^{-3}$~eV/\AA. Fully relaxed HSE06 structures obtained with the complete ``light'' numerical settings were additionally calculated to assess the effect of the exchange-correlation approximation on the geometry.

The surface test was performed for OER energetics using a stoichiometric monolayer slab model of the $\beta$-NiOOH(001) surface, which has previously been examined in first-principles studies of nickel oxyhydroxides \cite{eslamibidgoli2017surface}. The surface unit cell contained four NiOOH formula units. The out-of-plane cell-vector length was 50~\AA\ and included the vacuum region. A dipole correction was applied normal to the slab. The cell vectors were relaxed only with HSE06 and were subsequently kept fixed in all other calculations. Atomic positions were relaxed separately with PBE, regularized SCAN (rSCAN) \cite{bartok2019regularized}, HSE06, and the Perdew--Burke--Ernzerhof hybrid functional (PBE0) \cite{10.1063/1.478522}. The Brillouin zone was sampled using a $4\times4\times1$ mesh and a Gaussian occupation broadening width of 0.01~eV. The SCF thresholds were $10^{-3}$~eV for the sum of Hamiltonian eigenvalues, $10^{-5}$~eV for the total energy, and $10^{-3}$ for the charge density. During structural optimization, forces were evaluated with an SCF accuracy of $10^{-3}$~eV/\AA, and the BFGS optimization was terminated when the maximum force component was below $10^{-2}$~eV/\AA. Self-consistent ACBN0 calculations used the L\"{o}wdin projector and Petukhov mixing on the PBE- and rSCAN-relaxed structures. We were unable to obtain a complete OER sequence with SCAN because of substantial numerical instability, whereas the rSCAN calculations converged after careful adjustment of the Pulay-mixing parameters and selection of the initial magnetic moments.

Scalar-relativistic effects were included in all calculations using the atomic zero-order regular approximation (atomic ZORA) as implemented in FHI-aims \cite{blum2009ab}.

\section{Results and discussion}\label{sec:results}

\subsection{Computational cost}

The computational cost was assessed for six bulk transition metal oxides: Cr$_2$O$_3$, Cu$_2$O, CuO, MnO, NiO, and CoO. All calculations used the same computing cluster, the same number of CPU cores, and the FHI-aims ``light'' numerical settings, including the corresponding basis and integration grids. ACBN0@PBE with Petukhov mixing required up to 1.49 times the PBE time per iteration, while ACBN0@PBE--FLL required up to 1.39 times the PBE time. By comparison, one HSE06 iteration was 12.6--33.0 times as expensive as the corresponding ACBN0 iteration with Petukhov mixing. ACBN0 often, but not always, requires more iterations than the reference method because the electron density and the shell-resolved $\bar U$ and $\bar J$ parameters must converge together. In the current implementation, the number of iterations depends strongly on the system under study and on numerical choices such as the mixing settings and the initial magnetic moments used in the calculations. Improving the convergence of the coupled SCF--ACBN0 cycle remains a subject for future work.

\subsection{Electronic structure of semiconductors and insulators calculated with ACBN0}

ACBN0 improves the band gap predictions relative to the reference PBE and SCAN functionals for the majority of the test set. In the present chemically diverse benchmark, L\"{o}wdin orthogonalization provided stable projected occupations with both numerical settings and all three investigated double-counting treatments.

The deviations of ACBN0 band gap values from the experimental ones are comparable to those obtained from single-point hybrid-functional calculations. The mean absolute errors (MAEs) and mean absolute percentage errors (MAPEs) with respect to the experimental measurements reported in the literature are summarized in Table~\ref{mae_mape_wide}. Full data on the calculated band gaps are presented in SI Tables~S2 and S3. When more than one experimental band gap is reported for a material, the midpoint of the full span of the listed values and ranges is used as the reference in these error measures. For HgO, the reported values are represented by the interval 1.90--2.13~eV. FLL gives the lowest aggregated errors for this test set. ACBN0@SCAN--FLL gives MAPEs of 20.5\% and 25.8\% with the ``light'' and ``tight'' numerical settings, respectively. The corresponding ACBN0@PBE--FLL values are 28.3\% and 36.9\%. For the same PBE-relaxed geometries, single-point HSE06 gives MAPEs of 19.8\% and 20.4\% with the ``light'' and ``tight'' settings, respectively, while the fully relaxed HSE06/``light'' calculations give 20.1\%.
As expected from its potential form, AMF produces smaller corrections than FLL for the strongly localized materials in this test set. Petukhov mixing gives aggregate errors between those obtained with FLL and AMF, with MAPEs of 30.4\% for ACBN0@SCAN and 39.4\% for ACBN0@PBE using the ``light'' numerical settings. 

The ACBN0 results depend on the numerical settings. As shown in Table~\ref{mae_mape_wide}, the ``light'' settings give smaller average band gap errors than the ``tight'' settings for the present test set. As discussed above, when the projector contains one radial function from the minimal basis, additional radial and polarization functions change how electronic weight is distributed between the corrected subspace and the remaining basis functions. The resulting reduction in the occupations of the chosen localized subspaces propagates into the self-consistent Hubbard parameters. The smaller errors obtained here with the ``light'' settings are specific to this projector definition and do not imply that a smaller basis is generally more accurate. The projector definition and basis size should therefore be selected according to the material and the specific problem under consideration. Unless stated otherwise, the discussion below uses the FLL results obtained with the ``light'' numerical settings.

As described in Section~\ref{sec:projectors}, the Hubbard orbitals are selected from the minimal basis, while higher-tier functions remain outside the corrected subspace. The present comparison therefore examines how this projector construction depends on the orbital basis. A complete-basis population analysis would require the projector construction to account for all basis functions used to represent the Kohn--Sham states. To illustrate the distinction while controlling integration accuracy, SI Table~S10 compares three orbital bases for the transition metal oxide subset using identical ``tight'' grids and geometries: the ``light'' basis, the ``tight'' basis, and an expanded basis containing all second-tier metal functions and second- and third-tier O functions. Through the mechanism described in Section~\ref{sec:projectors}, the ``light'' basis assigns larger occupations to the selected subspaces and generally gives larger self-consistent Hubbard parameters and band gaps. The metal Mulliken moments do not vary monotonically across all materials because they include spin contributions from all atomic orbitals, including higher-tier functions outside the Hubbard subspace, whose redistribution upon basis expansion is material dependent. These tests quantify basis dependence for the chosen minimal-shell projector. They are not intended to establish a universal basis-convergence criterion.

To compare chemically distinct crystalline semiconductors, the test set was divided into $3d$ transition metal oxides, alkaline earth and post-transition metal oxides, and nitrides. For the $3d$ transition metal oxide subset (Table~\ref{mae_mape_tmo}), ACBN0@SCAN--FLL and ACBN0@PBE--FLL with the ``light'' numerical settings give MAEs of 0.63 and 0.76~eV and MAPEs of 24.1\% and 25.8\%, respectively. Depending on the numerical settings and geometry, HSE06 gives MAPEs of 19.1--23.8\% for this subset. For the combined alkaline earth and post-transition metal oxide subset (Table~\ref{mae_mape_aemo}), ACBN0@SCAN--FLL with the ``light'' numerical settings gives an MAE of 0.80~eV and a MAPE of 15.2\%, compared with HSE06 MAPEs of 19.2--21.7\%.

\begin{table*}[t]
\centering
\caption{Band gap mean absolute errors (MAEs) and mean absolute percentage errors (MAPEs) relative to the midpoint of the reported experimental span for each material in the complete test set. ``Relaxed'' denotes HSE06 calculations at HSE06-relaxed geometries, and ``sp'' denotes single-point calculations. All other results are single-point electronic-structure calculations at geometries relaxed with PBE using the ``tight'' numerical settings.}
\begin{tabular}{lcccc}
\toprule
\textbf{Method} & \multicolumn{2}{c}{\textbf{MAE (eV)}} & \multicolumn{2}{c}{\textbf{MAPE (\%)}} \\
\cmidrule(lr){2-3} \cmidrule(lr){4-5}
& \textbf{``light''} & \textbf{``tight''} & \textbf{``light''} & \textbf{``tight''} \\
\midrule
HSE06 (relaxed) & 0.74 & - & 20.1 & - \\
\hline
PBE & 2.05 & 2.04 & 55.3 & 55.0 \\
SCAN (sp) & 1.52 & 1.49 & 41.5 & 40.4 \\
HSE06 (sp) & 0.74 & 0.76 & 19.8 & 20.4 \\
\hline
\addlinespace[0.5ex]
ACBN0@PBE--Petukhov & 1.64 & 1.70 & 39.4 & 41.6 \\
ACBN0@SCAN--Petukhov & 1.26 & 1.27 & 30.4 & 30.3 \\
ACBN0@PBE--FLL & 1.13 & 1.48 & 28.3 & 36.9 \\
ACBN0@SCAN--FLL & 0.79 & 1.04 & 20.5 & 25.8 \\
ACBN0@PBE--AMF & 1.74 & 1.76 & 42.9 & 43.7 \\
ACBN0@SCAN--AMF & 1.29 & 1.31 & 30.6 & 31.2 \\
\bottomrule
\end{tabular}
\label{mae_mape_wide}
\end{table*}

\begin{table*}[t]
\centering
\caption{Band gap MAEs and MAPEs relative to the midpoint of the reported experimental span for each material in the $3d$ transition metal oxide subset. ``Relaxed'' denotes HSE06 calculations at HSE06-relaxed geometries, and ``sp'' denotes single-point calculations. All other results are single-point electronic-structure calculations at geometries relaxed with PBE using the ``tight'' numerical settings.}
\begin{tabular}{lcccc}
\toprule
\textbf{Method} & \multicolumn{2}{c}{\textbf{MAE (eV)}} & \multicolumn{2}{c}{\textbf{MAPE (\%)}} \\
\cmidrule(lr){2-3} \cmidrule(lr){4-5}
 & \textbf{``light''} & \textbf{``tight''} & \textbf{``light''} & \textbf{``tight''} \\
\midrule
HSE06 (relaxed) & 0.66 & - & 23.8 & - \\
\hline
PBE & 2.05 & 2.03 & 70.4 & 69.9 \\
SCAN (sp) & 1.46 & 1.38 & 53.5 & 51.2 \\
HSE06 (sp) & 0.52 & 0.56 & 19.1 & 20.7 \\
\hline
\addlinespace[0.5ex]
ACBN0@PBE--Petukhov & 1.13 & 1.27 & 34.5 & 39.4 \\
ACBN0@SCAN--Petukhov & 0.88 & 0.88 & 28.9 & 28.2 \\
ACBN0@PBE--FLL & 0.76 & 1.05 & 25.8 & 33.9 \\
ACBN0@SCAN--FLL & 0.63 & 0.70 & 24.1 & 24.8 \\
ACBN0@PBE--AMF & 1.36 & 1.40 & 42.8 & 44.4 \\
ACBN0@SCAN--AMF & 0.94 & 0.96 & 29.4 & 30.3 \\
\bottomrule
\end{tabular}
\label{mae_mape_tmo}
\end{table*}

\begin{table*}[t]
\centering
\caption{Band gap MAEs and MAPEs relative to the midpoint of the reported experimental span for each material in the alkaline earth and post-transition metal oxide subset. ``Relaxed'' denotes HSE06 calculations at HSE06-relaxed geometries, and ``sp'' denotes single-point calculations. All other results are single-point electronic-structure calculations at geometries relaxed with PBE using the ``tight'' numerical settings.}
\begin{tabular}{lcccc}
\toprule
\textbf{Method} & \multicolumn{2}{c}{\textbf{MAE (eV)}} & \multicolumn{2}{c}{\textbf{MAPE (\%)}} \\
\cmidrule(lr){2-3} \cmidrule(lr){4-5}
 & \textbf{``light''} & \textbf{``tight''} & \textbf{``light''} & \textbf{``tight''} \\
\midrule
HSE06 (relaxed) & 0.87 & - & 19.2 & - \\
\hline
PBE & 2.07 & 2.05 & 44.7 & 44.3 \\
SCAN (sp) & 1.60 & 1.59 & 32.9 & 32.5 \\
HSE06 (sp) & 0.96 & 0.95 & 21.7 & 21.4 \\
\hline
\addlinespace[0.5ex]
ACBN0@PBE--Petukhov & 2.04 & 2.03 & 43.1 & 43.2 \\
ACBN0@SCAN--Petukhov & 1.56 & 1.56 & 31.3 & 31.4 \\
ACBN0@PBE--FLL & 1.30 & 1.75 & 27.9 & 38.6 \\
ACBN0@SCAN--FLL & 0.80 & 1.26 & 15.2 & 25.3 \\
ACBN0@PBE--AMF & 2.04 & 2.03 & 43.2 & 43.2 \\
ACBN0@SCAN--AMF & 1.56 & 1.57 & 31.3 & 31.5 \\
\bottomrule
\end{tabular}
\label{mae_mape_aemo}
\end{table*}

The comparison of ACBN0 with PBE and SCAN references reveals systematic changes across the $3d$ transition metal series. For the early transition metal oxide TiO$_2$, ACBN0@SCAN gives band gaps closer to experiment than ACBN0@PBE, although both underestimate the experimental values. For rutile TiO$_2$ (experimental gap 3.05--3.065~eV), the calculated ACBN0@SCAN and ACBN0@PBE gaps are 2.71 and 2.28~eV, respectively. For anatase TiO$_2$ (experimental gap 3.3--3.4~eV), the ACBN0@SCAN and ACBN0@PBE gaps are 3.03 and 2.56~eV, respectively. The larger ACBN0@SCAN gaps are consistent with the stronger localization of Ti-$3d$ states produced by the SCAN parent functional.

Cr$_2$O$_3$ poses a challenge for both hybrid and DFT+$U$ methods because both can overlocalize the Cr-$3d$ states. In the present calculations, HSE06 gives band gaps of $4.46$--$4.54$~eV, depending on the numerical settings, compared with the experimental range of $3.0$--$3.2$~eV. Previous calculations have likewise shown that the predicted Cr$_2$O$_3$ band gap is sensitive to the fraction and screening of exact exchange, and that standard HSE06 can overestimate it \cite{navarrete2018hybrid,guo2012electronic}. ACBN0@PBE--FLL gives smaller gaps of 2.72 and 2.51~eV with the ``light'' and ``tight'' settings, respectively, whereas ACBN0@SCAN--FLL gives 3.88 and 3.76~eV. Thus, the PBE-based calculation underestimates the experimental gap, while the stronger localization inherited from SCAN produces an overestimation in the same manner as the hybrid functional.

For MnO, representing the middle of the $3d$ transition metal series in the present test set, ACBN0@SCAN gives a larger band gap that is closer to experiment than ACBN0@PBE. For an experimental gap of approximately $3.9$--$4.1$~eV, the calculated values are 2.70 and 2.01~eV, respectively. In this Mott--Hubbard insulator, the stronger localization produced by the SCAN parent functional increases the gap more effectively than PBE.

For late members of the $3d$ transition metal series, the trend is reversed, and ACBN0@SCAN tends to overestimate band gap values. For CoO, ACBN0@SCAN--FLL with the ``light'' numerical settings yields a gap of 3.21~eV (experimental value 2.6--2.8~eV). In contrast, ACBN0@PBE--FLL predicts a gap of 2.60~eV, at the lower bound of the experimental range. In the case of NiO, ACBN0@SCAN--FLL yields 4.41~eV compared with 3.67~eV from ACBN0@PBE--FLL (experimental value 4.0--4.3~eV).

CoO also illustrates spontaneous electronic symmetry breaking in calculations with self-consistent Hubbard parameters, indicating numerical sensitivity of the underlying meta-GGA calculation and the coupled ACBN0 update. In the ACBN0@SCAN--FLL solution, the two crystallographically equivalent Co sites acquire slightly different local magnetic moments but markedly different self-consistent Hubbard parameters, although the atomic structure is fixed. The stronger localization produced by SCAN may make the SCF cycle sensitive to small differences between the site-resolved occupation matrices, and the density-dependent ACBN0 update may then amplify this imbalance by assigning different Hubbard parameters to the two sites.

\begin{figure}[!t]
\centering
\includegraphics[width=0.96\columnwidth]{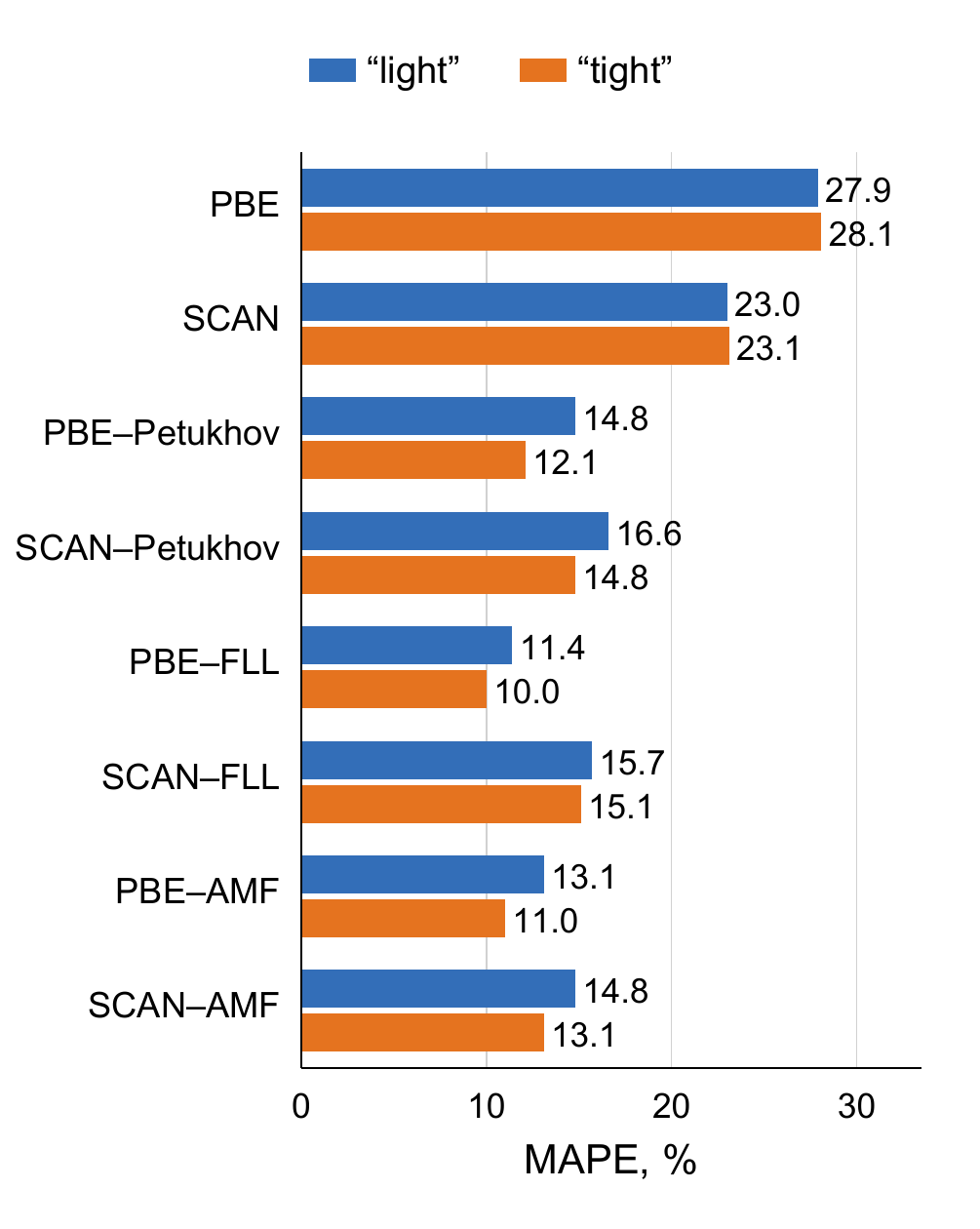}
\caption{MAPE of the calculated Mulliken magnetic moments relative to fully relaxed HSE06/``light''. The ``light'' and ``tight'' labels in the legend refer to the numerical settings. For the ACBN0 entries, the functional preceding the double-counting treatment denotes the parent functional. For the symmetry-broken ACBN0@SCAN--FLL solution of CoO, the mean magnitude of the two Co moments is used. The percentages use unrounded calculated and HSE06 reference moments; the moments in Supporting Information Tables~S4 and S5 are rounded for display.}
\label{fig:moment_hse_deviations}
\end{figure}

\begin{figure*}[t]
    \centering    \includegraphics[width=0.95\linewidth]{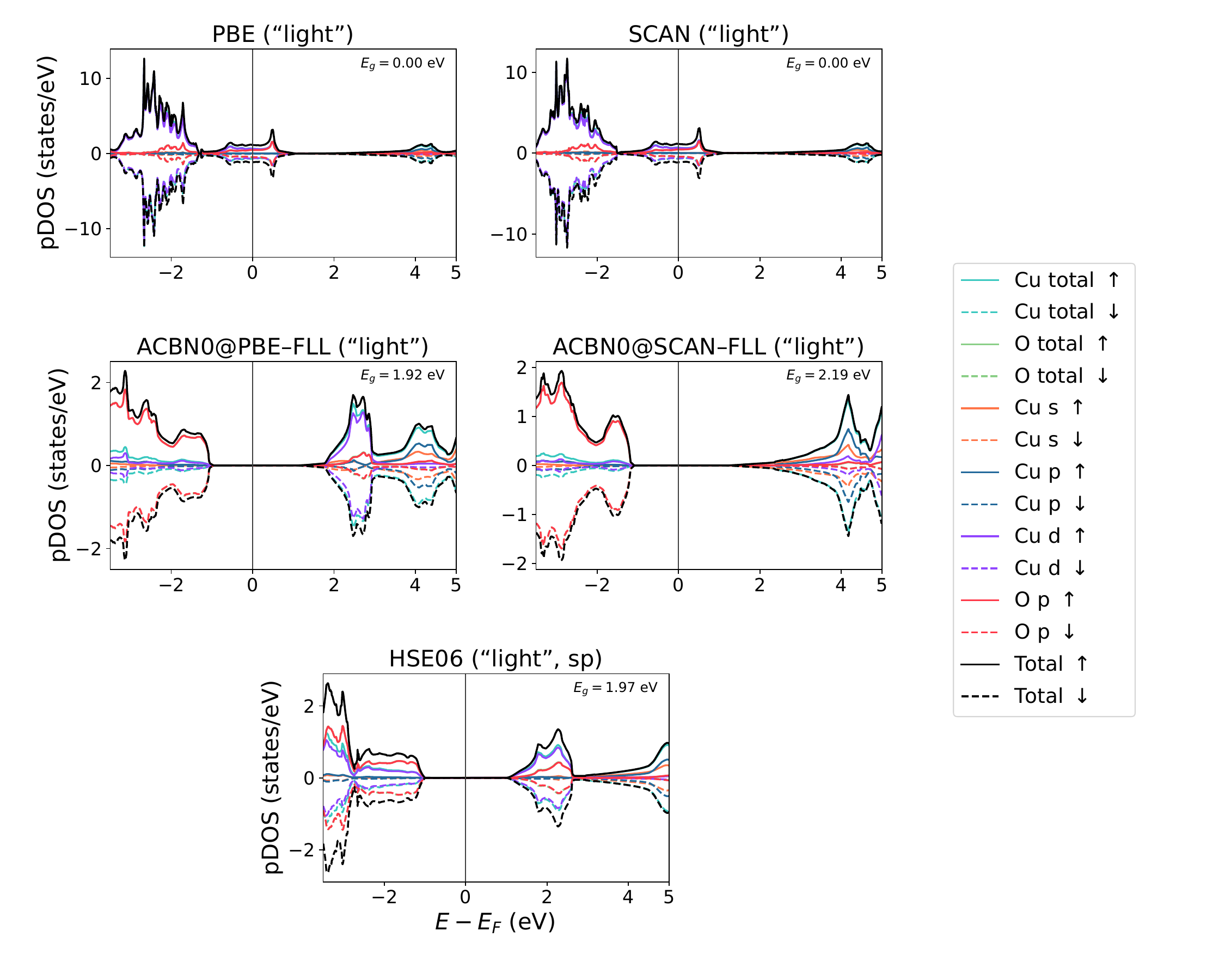}
    \caption{Projected densities of states (pDOS) for CuO calculated with PBE, SCAN, HSE06, ACBN0@PBE--FLL, and ACBN0@SCAN--FLL at the same PBE-relaxed geometry. All calculations used the ``light'' numerical settings. }
    \label{CuO}
\end{figure*}

\begin{figure*}[t]
    \centering    \includegraphics[width=0.95\linewidth]{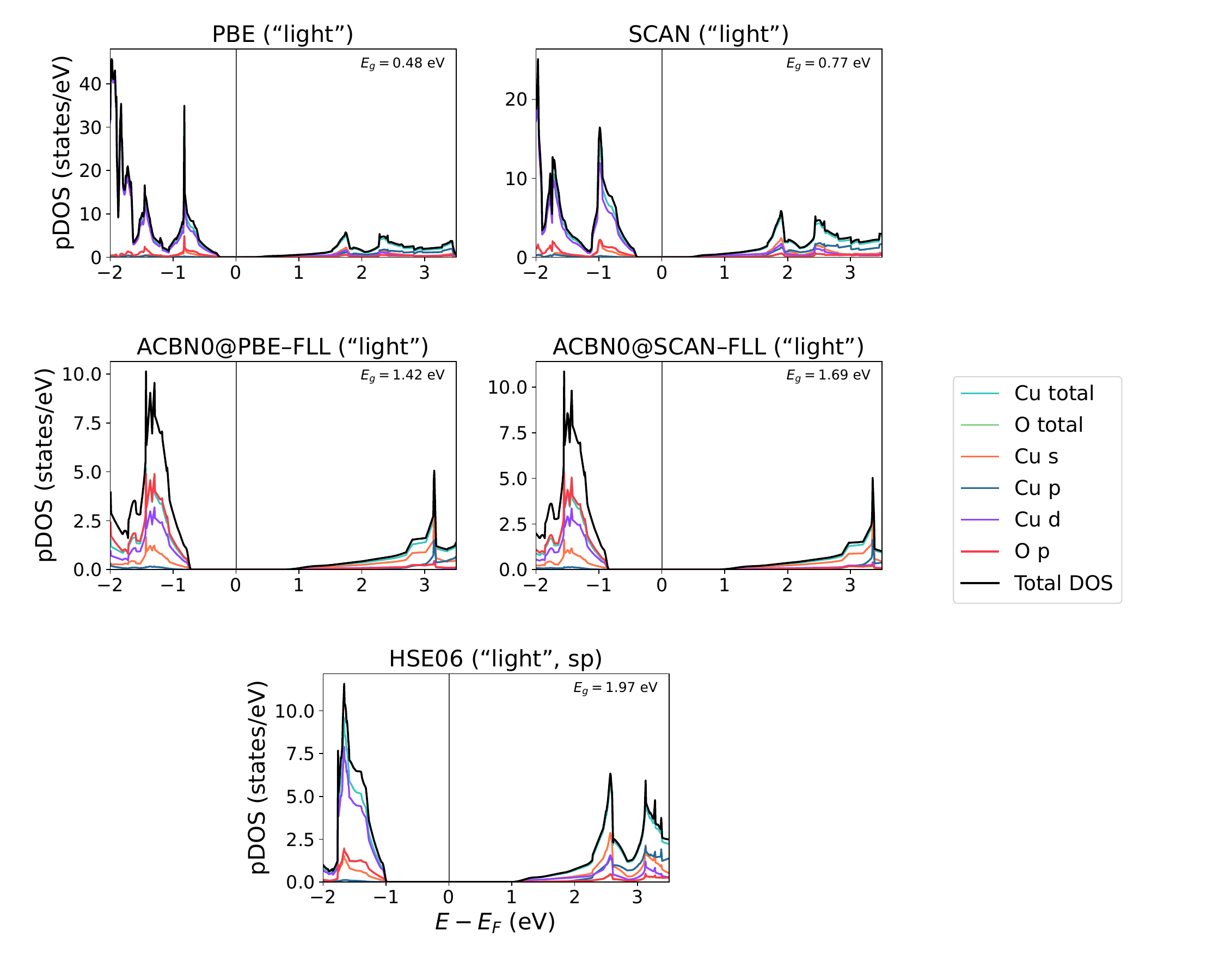}
    \caption{pDOS for Cu$_2$O calculated with PBE, SCAN, HSE06, ACBN0@PBE--FLL, and ACBN0@SCAN--FLL at the same PBE-relaxed geometry. All calculations used the ``light'' numerical settings. }
    \label{Cu2O}
\end{figure*}

The contrasting performance of ACBN0 for CuO and Cu$_2$O illustrates how the nature of the band gap determines the response to the Hubbard correction. CuO is a prototypical charge-transfer insulator (Cu$^{2+}$, $d^9$), where the band gap opens between the O-2$p$-dominated valence-band maximum (VBM) and the Cu-3$d$-dominated conduction-band minimum (CBM) (Fig.~\ref{CuO}) \cite{PhysRevB.38.11322}. In this system, both the $U$ correction in ACBN0 and exact exchange in HSE06 shift the unoccupied Cu-3$d$ states upward and the occupied O-2$p$ states downward, widening the gap. Consequently, ACBN0@PBE and ACBN0@SCAN yield gaps of 1.92 and 2.19~eV, respectively, compared with the experimental range of 1.35--1.40~eV \cite{koffyberg1982photoelectrochemical,Meyer2012}. Although the self-consistent $U_{\mathrm{Cu-3d}}$ values are similar for ACBN0@PBE--FLL (15.45~eV) and ACBN0@SCAN--FLL (15.91~eV) with the ``light'' numerical settings (Supporting Information Tables~S6 and S8), the CBM character differs. ACBN0@SCAN reinforces the localization already produced by SCAN. The upward shift of the unoccupied Cu-3$d$ states exposes Cu-4$s$- and Cu-4$p$-derived states near the conduction-band edge. ACBN0@PBE retains a Cu-3$d$/O-2$p$ contribution at the CBM, closer to HSE06 and consistent with the spectroscopic evidence for strong Cu-3$d$/O-2$p$ hybridization in CuO \cite{PhysRevB.38.11322}.

\begin{figure*}[t]
\centering
\includegraphics[width=0.95\linewidth]{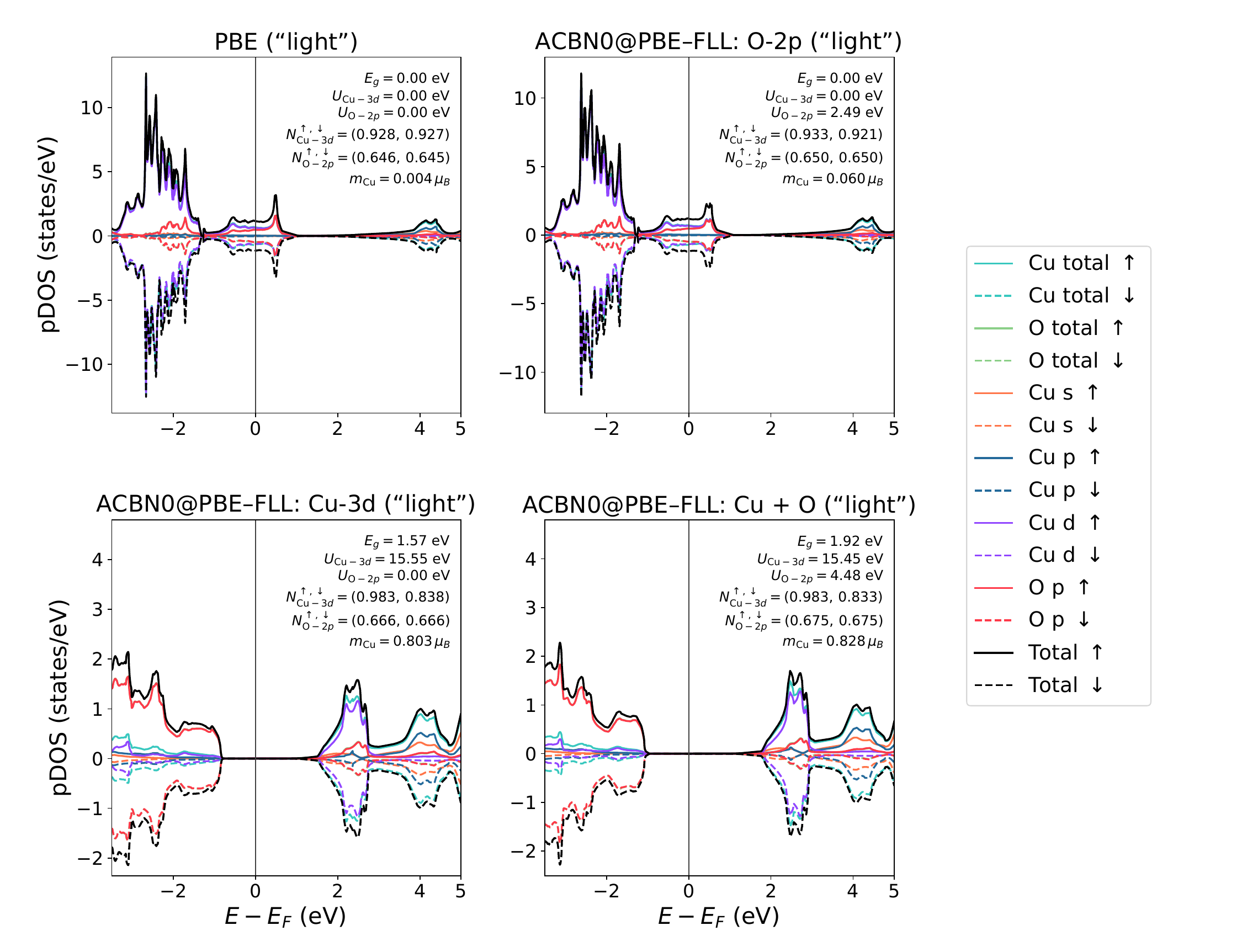}
\caption{Selective ACBN0@PBE--FLL corrections for CuO at the same PBE/``tight''-relaxed geometry with the ``light'' numerical settings: pure PBE, correction to O-$2p$ only, Cu-$3d$ only, and both shells. Each panel reports the band gap, effective Hubbard parameters, local Cu Mulliken moments, and L\"{o}wdin occupations for Cu-$3d$ and O-$2p$. $N^{\sigma}$ denotes the mean occupation over the $2l+1$ orbital components of the specified shell in spin channel $\sigma$.}
\label{fig:CuO_selective_U}
\end{figure*}

Selective ACBN0@PBE--FLL calculations at the same PBE/``tight''-relaxed geometry and with identical ``light'' numerical settings isolate the roles of Cu-$3d$ and O-$2p$ (Fig.~\ref{fig:CuO_selective_U}). Pure PBE and the O-only correction ($U_{\mathrm{O-2p}}=2.49$~eV) remain metallic and almost nonmagnetic: the Cu-$3d$ occupations per orbital in the two spin channels change only from $(0.928,0.927)$ to $(0.933,0.921)$, while the O-$2p$ occupations remain approximately 0.65 in both channels. Correcting Cu-$3d$ alone gives $U_{\mathrm{Cu-3d}}=15.55$~eV, separates the Cu occupations to $(0.983,0.838)$, and stabilizes an AFM insulating state with a gap of 1.566~eV and a Cu moment of $0.803~\mu_B$. The O occupations increase to $(0.666,0.666)$. With both shells corrected, $U_{\mathrm{Cu-3d}}=15.45$~eV and $U_{\mathrm{O-2p}}=4.48$~eV, the Cu and O occupations become $(0.983,0.833)$ and $(0.675,0.675)$, respectively, and the gap increases to 1.920~eV while the Cu moment changes only slightly, to $0.828~\mu_B$. The pDOS associates the main conduction-band reorganization with the Cu correction. Adding the O correction shifts occupied O-$2p$ states downward without substantially changing the conduction-band profile. Thus, correcting Cu-$3d$ is essential for the magnetic insulating solution, whereas correcting O-$2p$ alone is insufficient and, in the combined calculation, adds 0.354~eV to an already overestimated gap. The increase in $U_{\mathrm{O-2p}}$ from 2.49 to 4.48~eV also shows that the two corrections are coupled through the self-consistent electronic state.

In contrast, Cu$_2$O is a conventional semiconductor with a filled Cu-$3d$ shell. Its valence band has hybridized Cu-$3d$/O-$2p$ character, whereas the CBM is Cu-$4s$-derived (Fig.~\ref{Cu2O}). The Cu-$4s$ contribution to the occupied Kohn--Sham states is weak, as reflected by its low average L\"{o}wdin occupation, $N^{\sigma}\approx0.19$, in each spin channel. This weak localized contribution strongly suppresses the corresponding renormalized density matrix elements, and ACBN0, applied to the Cu-$4s$ shell, consequently gives $U_{\mathrm{Cu-4s}}\sim0.1$~eV. The $+U$ correction to the Cu-$4s$ shell is therefore negligible. Correcting the occupied Cu-$3d$ and O-$2p$ states increases the gap from 0.48~eV with PBE to 1.42~eV with ACBN0@PBE, while the Cu-$4s$ conduction edge remains too low. ACBN0@SCAN gives 1.69~eV; both values remain below the experimental gap of approximately 2.4~eV.

Overall, the combination of meta-GGA SCAN with a Hubbard correction, whether evaluated by ACBN0 or with fixed $U$, should be assessed for each material and target property. The stronger localization produced by SCAN can complement the Hubbard correction, as observed for some early and mid-series transition metal oxides in this benchmark, but the combined effect can also overlocalize electronic states. In the present results, this limitation is most apparent for late transition metal oxides with nearly filled $d$ shells, including CoO, NiO, and CuO.
 
For the alkaline earth and post-transition metal oxides, Petukhov mixing gives results that are numerically close to the AMF limit and therefore produces smaller increases in the band gaps than FLL. In these cases, the occupation matrix within a corrected shell is close to $n^{\sigma}_{mm'}=N^{\sigma}\delta_{mm'}$. The small deviations from this uniform occupation reduce the Petukhov mixing factor toward $\alpha=0$ [Eq.~(\ref{mixing2})], irrespective of whether the projected occupations are close to unity. Specifically, the AMF expression for the potential component is
\begin{equation}
    V^{\sigma}_{mm'}(\alpha^{I,n,l}=0)=-U^{I,n,l}\left(n_{mm'}^{\sigma}-N^{\sigma}\delta_{mm'}\right),
\end{equation}
whereas the FLL expression is
\begin{equation}
    V^{\sigma}_{mm'}(\alpha^{I,n,l}=1)=-U^{I,n,l}\left(n_{mm'}^{\sigma}-\frac{1}{2}\delta_{mm'}\right).
\end{equation}
When $n^{\sigma}_{mm'}\approx N^{\sigma}\delta_{mm'}$, the AMF potential is small, while the FLL potential retains the shell-wide shift $-U^{I,n,l}(N^{\sigma}-\tfrac{1}{2})\delta_{mm'}$. Thus, near-uniform occupations suppress the AMF correction, while unless $N^{\sigma}$ is close to one half, FLL can still produce a substantial shift and a larger change in the band gap, as observed for the closed-shell materials examined here.

For Bi$_2$O$_3$, applying ACBN0 to O-$2p$ alone or additionally to combinations of Bi-$6s$, Bi-$6p$, and deep semicore Bi-$5d$ shells changes the band gap by no more than 0.02~eV. The VBM has strongly hybridized Bi-$6s$/O-$2p$ character, whereas the CBM is dominated by Bi-$6p$. The self-consistent Hubbard parameters for the Bi-$6s$ and Bi-$6p$ band-edge contributions are close to zero, while the larger Bi-$5d$ correction acts on deep semicore states. Adding the Bi shells therefore has little effect on the fundamental gap relative to correcting O-$2p$ alone.

For AlN, ACBN0 produces little improvement over PBE or SCAN because the self-consistent correction for the relevant Al-$3p$ shell is only about 0.1~eV (Supporting Information Tables~S6 and S8). In GaN, the VBM is dominated by N-$2p$ states and the CBM by Ga-$4s$ states, with a smaller Ga-$4p$ contribution. With the ``light'' settings, the N-$2p$ occupations are approximately 0.51 per orbital and spin and nearly equal across the three components: the FLL shift is small because the occupations are close to one half (see Eq.~\eqref{V}), and the AMF-like shift is small because their deviations from the shell average are small. The localized Ga-$3d$ semicore shell instead acquires large Hubbard parameters, 24.56 and 24.48~eV for ACBN0@PBE--FLL and ACBN0@SCAN--FLL, respectively. All energies in the following comparison are referenced to the internal energy zero of periodic FHI-aims. The Ga-$3d$ band center shifts from $-20.38$~eV with PBE to $-32.49$~eV with ACBN0@PBE--FLL and from $-21.02$~eV with SCAN to $-33.13$~eV with ACBN0@SCAN--FLL. Relative to PBE and SCAN, the corresponding VBM shifts are $-0.152$ and $-0.161$~eV, while the CBM shifts are only $+0.013$ and $+0.012$~eV. The gaps consequently increase from 1.76~eV with PBE to 1.92~eV with ACBN0@PBE--FLL and from 2.06~eV with SCAN to 2.23~eV with ACBN0@SCAN--FLL. The semicore correction affects the band edges indirectly through reduced $p$--$d$ interaction, while the direct correction to the N-$2p$ valence edge remains small.

\begin{table*}[t]
\centering
\caption{Aggregate errors of the Mulliken magnetic moments relative to experiment for the antiferromagnetic transition metal oxides. The MAE and MAPE use the midpoint of the full span of the experimental values listed for each material in Supporting Information Table~S4. A single reported value is used unchanged. For the symmetry-broken ACBN0@SCAN--FLL solution of CoO, the mean magnitude of the two Co moments is used. ``Relaxed'' denotes HSE06 calculations on HSE06-relaxed geometries, and ``sp'' denotes single-point calculations.}
\setlength{\tabcolsep}{4pt}
\begin{tabular}{lcccc}
\toprule
\textbf{Method} & \multicolumn{2}{c}{\textbf{Experimental MAE ($\mu_B$)}} & \multicolumn{2}{c}{\textbf{Experimental MAPE (\%)}} \\
\cmidrule(lr){2-3} \cmidrule(lr){4-5}
 & \textbf{``light''} & \textbf{``tight''} & \textbf{``light''} & \textbf{``tight''} \\
\midrule
HSE06 (relaxed) & 0.42 & - & 14.4 & - \\
\hline
PBE & 0.67 & 0.67 & 37.0 & 37.0 \\
SCAN (sp) & 0.58 & 0.57 & 33.7 & 33.0 \\
HSE06 (sp) & 0.42 & 0.42 & 14.5 & 14.6 \\
\hline
\addlinespace[0.5ex]
ACBN0@PBE--Petukhov & 0.40 & 0.41 & 16.5 & 16.7 \\
ACBN0@SCAN--Petukhov & 0.37 & 0.38 & 16.6 & 16.9 \\
ACBN0@PBE--FLL & 0.36 & 0.39 & 13.9 & 14.8 \\
ACBN0@SCAN--FLL & 0.35 & 0.35 & 16.0 & 15.6 \\
ACBN0@PBE--AMF & 0.42 & 0.42 & 16.4 & 15.7 \\
ACBN0@SCAN--AMF & 0.38 & 0.38 & 16.7 & 15.9 \\
\bottomrule
\end{tabular}
\label{mae_mape_mag}
\end{table*}

The magnetic moments provide a complementary measure of electron localization. Using the midpoint of the reported experimental span for each of the five antiferromagnetically ordered oxides, as in the band gap analysis, PBE gives an aggregate MAPE of 37.0\% with both numerical settings, and SCAN reduces it to 33.0--33.7\% (Table~\ref{mae_mape_mag}). Every ACBN0 variant lowers the MAPE to 13.9--16.9\% and the MAE to 0.35--0.42~$\mu_B$, comparable to the HSE06 values of 14.4--14.6\% and 0.42~$\mu_B$. FLL gives the lowest MAPE for both parent functionals with both numerical settings, reaching 13.9\% for ACBN0@PBE with the ``light'' settings. For SCAN, FLL also gives the lowest MAE, 0.35~$\mu_B$ with both settings.

This comparison with experiment must be interpreted with care. The calculated quantities are spin-only Mulliken moments from collinear calculations, whereas experimental values can depend on the spatial definition of the atomic moment and can include an orbital contribution. Figure~\ref{fig:moment_hse_deviations} therefore compares the calculated moments with results obtained after full HSE06 geometry relaxation with the ``light'' settings. This reference compares the complete computational protocols, including the effect of the relaxed geometry. The PBE MAPE is 27.9\% and 28.1\% for the ``light'' and ``tight'' settings, respectively, and decreases to 10.0--14.8\% with ACBN0@PBE. The corresponding SCAN values decrease from 23.0--23.1\% to 13.1--16.6\% with ACBN0@SCAN. FLL gives the lowest errors among the PBE-based calculations, reaching 11.4\% and 10.0\% with the ``light'' and ``tight'' settings. The ACBN0@SCAN errors are generally slightly larger than those of ACBN0@PBE. This behavior is consistent with the band gap trends discussed above: SCAN already localizes the spin density more strongly than PBE, and the additional Hubbard correction can produce excessive localization in some materials. Nevertheless, all investigated ACBN0 variants substantially reduce the aggregate deviation from the HSE06-relaxed reference relative to their uncorrected parent functionals.

The material-resolved results are presented in Supporting Information Tables~S4 and S5. In MnO and NiO, ACBN0 increases the metal moments toward the HSE06 and experimental references, although the stronger corrections can produce a moderate overestimate. In CuO, PBE and SCAN converge to nearly vanishing site moments, while all three ACBN0 double-counting treatments stabilize spin-polarized solutions. ACBN0 therefore corrects the pronounced underlocalization in these compounds, with the remaining deviations reflecting the balance between restoration of local moments and excessive localization. Cr$_2$O$_3$ and CoO show that stronger localization does not necessarily improve agreement with experimental local moments. In Cr$_2$O$_3$, PBE already overestimates the reported ordered moments, and both HSE06 and ACBN0@PBE--FLL increase the Cr moments further. This trend is consistent with reduced Cr-$3d$/O-$2p$ spin delocalization, whereas the experimental reduction from the ionic limit has been attributed to covalency and zero-point spin fluctuations \cite{brown2002determination,skovhus2022magnons}. In CoO, ACBN0@PBE--FLL increases the underestimated PBE moments toward the HSE06 values, although the calculated spin-only Mulliken moments remain below experiment. The ACBN0@SCAN--FLL solution also breaks the equivalence of the two Co sites, as discussed above.

When band gaps and magnetic moments are considered together, FLL is the most consistent of the three double-counting treatments for this test set, although the preferred parent functional remains material dependent.

\subsection{Adsorption on the (001) surface of $\beta$-nickel oxyhydroxide}

To evaluate the method for a low-dimensional system, we investigated the OER energetics on the $\beta$-NiOOH(001) surface.

The four-electron OER process was analyzed using the alkaline OER mechanism described in Ref.~\cite{Liang_2021}:

\begin{align}
\text{*} + \text{OH}^- &= \text{*OH} + e^-, \label{eq:oer_step1}\\
\text{*OH} + \text{OH}^- &= \text{*O} + \text{H}_2\text{O}_{(\text{l})} + e^-, \label{eq:oer_step2}\\
\text{*O} + \text{OH}^- &= \text{*OOH} + e^-, \label{eq:oer_step3}\\
\text{*OOH} + \text{OH}^- &= \text{*} + \text{O}_{2(\text{g})} + \text{H}_2\text{O}_{(\text{l})} + e^-. \label{eq:oer_step4}
\end{align}

To estimate the energetics of reactions~\eqref{eq:oer_step1}--\eqref{eq:oer_step4}, the following equations were used, where $U_{\mathrm{RHE}}$ is the applied electrode potential relative to the reversible hydrogen electrode (RHE) and $e$ is the positive elementary charge:
\begin{align}
\Delta E_1^{\mathrm{el}} &= E(\mathrm{*OH}) - E(\mathrm{*}) + \tfrac{1}{2}E(\mathrm{H}_2) \notag\\
&\quad - E(\mathrm{H}_2\mathrm{O}) - eU_{\mathrm{RHE}}, \label{eq:E1el}\\
\Delta E_2^{\mathrm{el}} &= E(\mathrm{*O}) - E(\mathrm{*OH}) \notag\\
&\quad + \tfrac{1}{2}E(\mathrm{H}_2) - eU_{\mathrm{RHE}}, \label{eq:E2el}\\
\Delta E_3^{\mathrm{el}} &= E(\mathrm{*OOH}) - E(\mathrm{*O}) + \tfrac{1}{2}E(\mathrm{H}_2) \notag\\
&\quad - E(\mathrm{H}_2\mathrm{O}) - eU_{\mathrm{RHE}}, \label{eq:E3el}\\
\Delta E_4^{\mathrm{el}} &= E(\mathrm{*}) - E(\mathrm{*OOH}) + \tfrac{1}{2}E(\mathrm{H}_2) \notag\\
&\quad + E(\mathrm{O}_2) - eU_{\mathrm{RHE}}. \label{eq:E4el}
\end{align}
Only DFT total electronic energies were considered; zero-point energy, entropic, and finite-temperature corrections were omitted. Although the formal reactions indicate liquid water and gaseous oxygen, the DFT energy expressions use isolated-molecule energies for H$_2$, H$_2$O, and O$_2$.
The calculated values of $\Delta E_i^{\mathrm{el}}$ are summarized in Table~\ref{thermodynamics}.

For the descriptors below, the reaction-step energies are evaluated at $U_{\mathrm{RHE}}=0$~V. The electronic-energy-derived overpotential, $\eta_{\mathrm{theory}}^{\mathrm{el}}$, was then estimated from the limiting reaction-step potential:
\begin{equation}\label{eq:limiting_potential}
    U_L^{\mathrm{el}}=\frac{1}{e}\max\left(\Delta E_1^{\mathrm{el}}, \Delta E_2^{\mathrm{el}},
\Delta E_3^{\mathrm{el}}, \Delta E_4^{\mathrm{el}}\right),
\end{equation}
\begin{equation}
\eta_{\mathrm{theory}}^{\mathrm{el}} =
U_L^{\mathrm{el}}- \frac{E_0^{\mathrm{el}}}{e},
\label{eq:overpotential}
\end{equation}
where $E_0^{\mathrm{el}}$ is calculated as
\begin{equation}
E_0^{\mathrm{el}} =
\frac{\Delta E_1^{\mathrm{el}} + \Delta E_2^{\mathrm{el}} +
\Delta E_3^{\mathrm{el}} + \Delta E_4^{\mathrm{el}}}{4}.
\label{eq:E0}
\end{equation}

The superscript ``el'' distinguishes these quantities from conventional free-energy OER descriptors. A free-energy-based overpotential is obtained from reaction-step free energies and referenced to the thermodynamic equilibrium potential, whereas Eqs.~\eqref{eq:limiting_potential}--\eqref{eq:E0} use only total electronic energies. The present descriptors are therefore used only to compare the relative performance of the semi-local, ACBN0, and hybrid functionals, without comparison to an experimental overpotential. Because zero-point-energy and entropic corrections are intermediate dependent, they could in principle change the potential-limiting step. Establishing those effects would require a separate free-energy analysis and is outside the scope of this benchmark.

The adsorption energies, $E_{\mathrm{ads}}$, were calculated using
\begin{equation}
E_{\mathrm{ads}} = E_{\mathrm{clean+ads}} - E_{\mathrm{clean}} - E_{\mathrm{adsorbate}},
\label{eq:Eads}
\end{equation}
where $E_{\mathrm{clean+ads}}$, $E_{\mathrm{clean}}$, and $E_{\mathrm{adsorbate}}$ are the total energies of the adsorbate-covered slab, the clean slab, and the molecular-reference energy for the adsorbate, respectively. The reference energies are $\tfrac{1}{2}E(\mathrm{O}_2)$ for O, $\tfrac{1}{2}E(\mathrm{H}_2)$ for H, and $\tfrac{1}{2}[E(\mathrm{H}_2)+E(\mathrm{O}_2)]$ for OH.

\begin{table*}[htbp]
\centering
\caption{Adsorption energies, $E_{\mathrm{ads}}$, calculated using different exchange--correlation functionals and numerical settings. Negative values correspond to exothermic adsorption. The energies are rounded to $10^{-3}$~eV.}
\label{adsorption_energies}
\begin{tabular}{llccc}
\toprule
Method & Numerical settings & \multicolumn{3}{c}{$E_{\mathrm{ads}}$, eV} \\
\cmidrule(lr){3-5}
       &           & O      & OH     & H      \\
\midrule
HSE06  & ``light''     & 1.017  & -0.549 & -1.402 \\
       & ``tight''     & 1.026  & -0.510 & -1.370 \\
PBE0   & ``light''     & 0.879  & -0.562 & -1.422 \\
       & ``tight''     & 0.889  & -0.520 & -1.384 \\
\midrule
PBE    & ``tight''     & 0.911  & -0.070 & -0.656 \\
rSCAN  & ``tight''     & 1.108  & -0.251 & -0.929 \\
\midrule
ACBN0@PBE (PBE slab)   & ``light'' & 1.060 & -0.126 & -0.692 \\
                       & ``tight'' & 0.961 & -0.088 & -0.665 \\
ACBN0@rSCAN (rSCAN slab) & ``light'' & 1.125 & -0.685 & -1.155 \\
                         & ``tight'' & 1.194 & -0.326 & -0.947 \\
\bottomrule
\end{tabular}
\end{table*}

\begin{table*}[htbp]
\centering
\caption{OER reaction-step energies, $\Delta E_i^{\mathrm{el}}$, calculated at $U_{\mathrm{RHE}}=0$~V using different exchange-correlation functionals and numerical settings. The energies are rounded to $10^{-3}$~eV.}
\label{thermodynamics}
\begin{tabular}{llcccc}
\toprule
Method & Numerical settings & \multicolumn{4}{c}{$\Delta E_i^{\mathrm{el}}$, eV} \\
\cmidrule(lr){3-6}
       &           & $\Delta E_1^{\mathrm{el}}$ & $\Delta E_2^{\mathrm{el}}$ & $\Delta E_3^{\mathrm{el}}$ & $\Delta E_4^{\mathrm{el}}$ \\
\midrule
HSE06  & ``light''     & 2.083  & 1.566  & 1.614  & 0.001   \\
       & ``tight''     & 2.106  & 1.536  & 1.608  & -0.018  \\
PBE0   & ``light''     & 2.075  & 1.441  & 1.754  & 0.004  \\
       & ``tight''     & 2.100  & 1.409  & 1.745  & -0.015 \\
\midrule
PBE    & ``tight''     & 2.430  & 0.981  & 0.737  & 0.852   \\
rSCAN  & ``tight''     & 2.327  & 1.359  & 1.850  & -0.381  \\
\midrule
ACBN0@PBE (PBE slab)   & ``light'' & 2.367 & 1.186 & 0.314 & 1.119 \\
                       & ``tight'' & 2.417 & 1.049 & 0.564 & 0.979 \\
ACBN0@rSCAN (rSCAN slab) & ``light'' & 1.927 & 1.809 & 1.333 & 0.153 \\
                         & ``tight'' & 2.256 & 1.519 & 1.638 & -0.250 \\
\bottomrule
\end{tabular}
\end{table*}

\begin{table*}[htbp]
\centering
\caption{OER descriptors obtained from the reaction-step energies: the mean electronic reaction-step energy, limiting potential, and electronic-energy-derived overpotential defined in Eqs.~\eqref{eq:E0}, \eqref{eq:limiting_potential}, and \eqref{eq:overpotential}, respectively.}
\label{tab:oer_summary}
\setlength{\tabcolsep}{10pt}
\begin{tabular}{llccc}
\toprule
Method & Numerical settings & $E_0^{\mathrm{el}}$, eV & $U_L^{\mathrm{el}}$, V & $\eta_{\mathrm{theory}}^{\mathrm{el}}$, V \\
\midrule
HSE06 & ``light'' & 1.316 & 2.083 & 0.767 \\
      & ``tight'' & 1.308 & 2.106 & 0.798 \\
PBE0  & ``light'' & 1.319 & 2.075 & 0.756 \\
      & ``tight'' & 1.310 & 2.100 & 0.790 \\
\midrule
PBE   & ``tight'' & 1.250 & 2.430 & 1.180 \\
rSCAN & ``tight'' & 1.289 & 2.327 & 1.038 \\
\midrule
ACBN0@PBE (PBE slab)   & ``light'' & 1.246 & 2.367 & 1.121 \\
                       & ``tight'' & 1.252 & 2.417 & 1.165 \\
ACBN0@rSCAN (rSCAN slab) & ``light'' & 1.306 & 1.927 & 0.621 \\
                         & ``tight'' & 1.291 & 2.256 & 0.965 \\
\bottomrule
\end{tabular}
\end{table*}

The optimized geometries provide an initial measure of the functional dependence of adsorption on the $\beta$-NiOOH(001) surface. The clean surface and the $\mathrm{*H}$, $\mathrm{*O}$, and $\mathrm{*OH}$ intermediates retain qualitatively similar local motifs under PBE, rSCAN, PBE0, and HSE06; the selected bond lengths show only moderate variations (Supporting Information Table~S12). The principal qualitative difference occurs for the $\mathrm{*OOH}$-related intermediate.

\begin{figure}[t]
\centering
\includegraphics[width=\columnwidth]{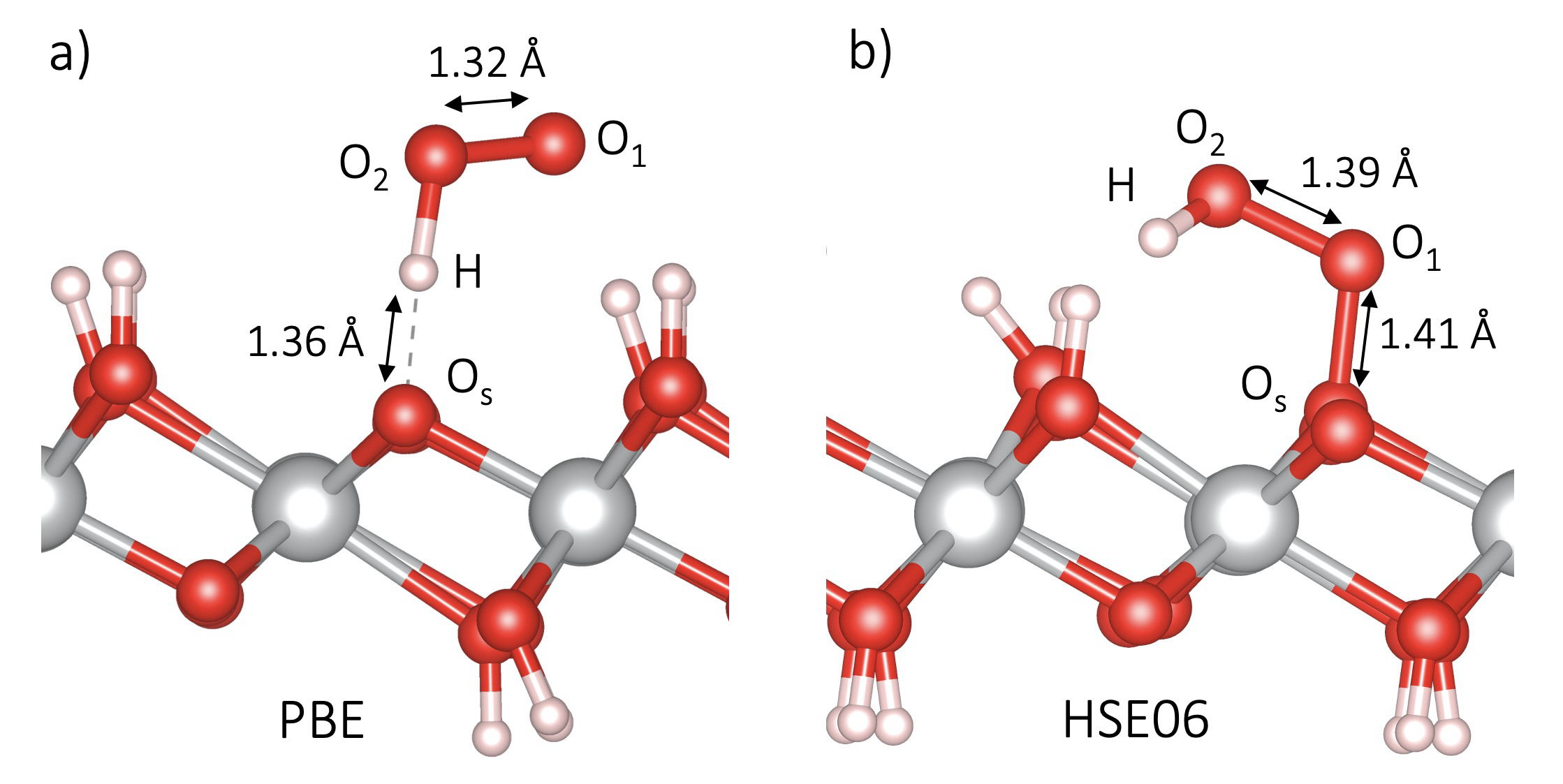}
\caption{Local geometry of the OOH-related intermediate in the monolayer slab model of the $\beta$-NiOOH(001) surface predicted by (a) PBE and (b) HSE06. Selected interatomic distances are indicated in \AA.}
\label{fig:ooh_geometry_comparison}
\end{figure}

As shown in Figure~\ref{fig:ooh_geometry_comparison}, HSE06 stabilizes an intact surface-bound $O_{\mathrm{s}}$--$O_1$--$O_2$ motif. PBE0 and rSCAN produce the same qualitative topology, with $O_{\mathrm{s}}$--$O_1$ and $O_1$--$O_2$ distances in the range expected for O--O bonding. PBE instead displaces the OOH fragment from the reactive surface oxygen site and stabilizes an H--$O_{\mathrm{s}}$ bond. This structural difference is essential for interpreting the OER energetics: the anomalously small $\Delta E_3^{\mathrm{el}}$ values obtained with PBE and ACBN0@PBE originate from the different PBE-relaxed $\mathrm{*OOH}$ topology described above. This result is consistent with observations in the literature that semi-local functionals can spuriously delocalize electron and hole polarons, while oxygen-hole formation and enhanced metal--oxygen covalency can promote oxygen oxidation and O--O bond formation in transition metal oxides \cite{rana2022delocalization,grimaud2017activating}. Although the present calculations do not include a direct charge-density or polaron analysis, the structural results are consistent with these mechanisms and illustrate a limitation of a single-point DFT+$U$ workflow.

Table~\ref{adsorption_energies} shows that the exchange--correlation treatment affects the adsorption energies of H, O, and OH differently, even though their adsorption geometries remain similar across the functionals. PBE substantially underbinds OH and H relative to both hybrid functionals, whereas its O adsorption energy lies within the HSE06--PBE0 interval. rSCAN strengthens OH and H adsorption and therefore reduces the discrepancy with the hybrid results. The response to ACBN0 depends in turn on the parent functional and the adsorbed state. With the ``tight'' settings, ACBN0@PBE changes all three adsorption energies by at most 0.050~eV and ACBN0@rSCAN by at most 0.086~eV. Both corrections slightly strengthen OH and H adsorption but make O adsorption less favorable. These changes reflect the state-specific redistribution of the Ni-$3d$ and O-$2p$ occupations upon adsorption. The considerably larger shifts obtained with the ``light'' ACBN0@rSCAN calculations, most clearly for OH and H, demonstrate that this redistribution is sensitive to the radial flexibility of the basis used together with the L\"owdin projector.

\begin{figure*}[t!]
\centering
\includegraphics[width=\textwidth]{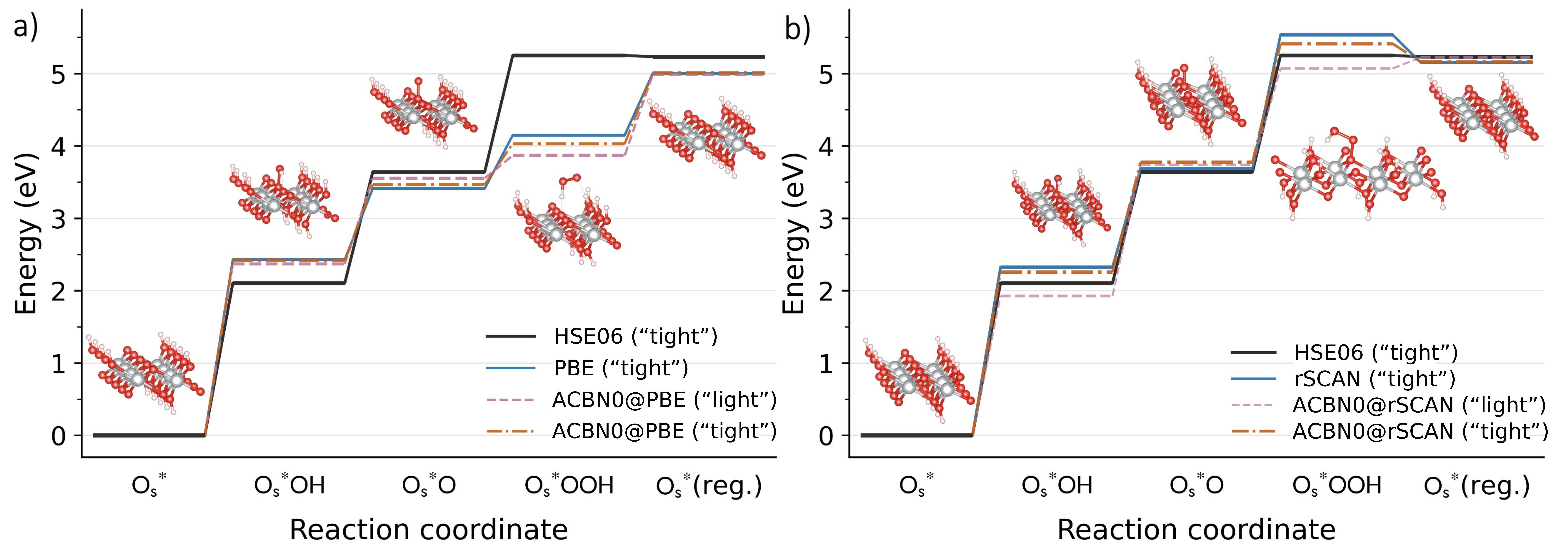}
\caption{Electronic-energy profiles for the four-electron OER cycle on the $\beta$-NiOOH(001) surface at $U_{\mathrm{RHE}}=0$~V. Panel (a) compares HSE06/``tight'' and PBE/``tight'' with ACBN0@PBE obtained using the ``light'' and ``tight'' numerical settings. Panel (b) compares HSE06/``tight'' and rSCAN/``tight'' with ACBN0@rSCAN obtained using the ``light'' and ``tight'' numerical settings. The profiles are cumulative sums of the reaction-step energies shown in Table~\ref{thermodynamics}. The structural insets show the clean and adsorbate-covered surfaces; red, gray, and white spheres denote O, Ni, and H atoms, respectively.}
\label{fig:oer_energy_profiles}
\end{figure*}

\begin{figure*}[t!]
\centering
\includegraphics[width=\textwidth]{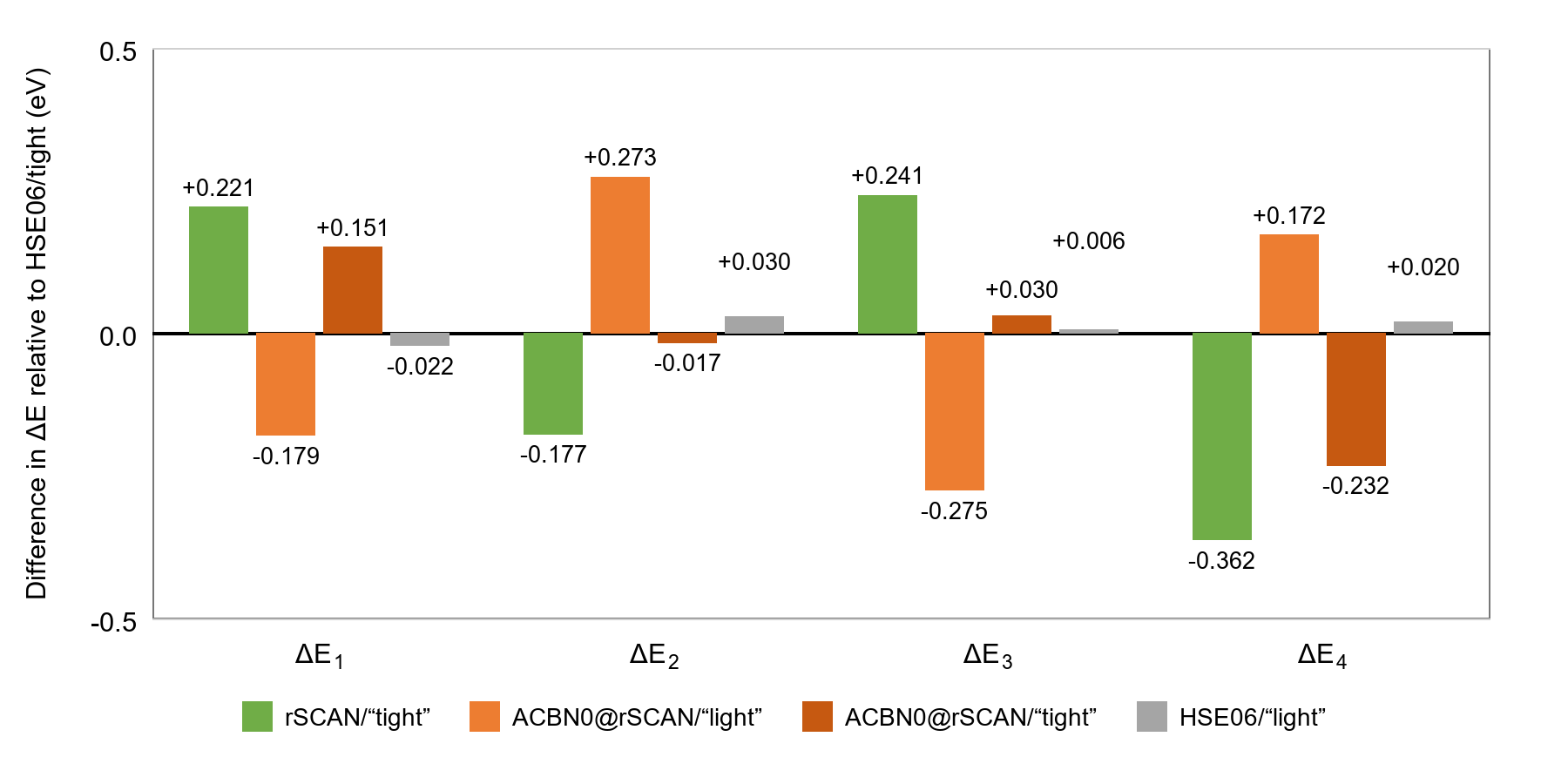}
\caption{Differences between the electronic OER reaction-step energies calculated with rSCAN, ACBN0@rSCAN, and HSE06 for the $\beta$-NiOOH(001) surface. For each step, the HSE06 value obtained with the ``tight'' numerical settings is taken as zero. rSCAN and ACBN0@rSCAN use rSCAN-relaxed geometries, whereas HSE06 uses HSE06-relaxed geometries. The labels give the signed deviations from HSE06/``tight'' in eV.}
\label{fig:oer_step_deviations}
\end{figure*}

Figure~\ref{fig:oer_energy_profiles} shows the complete cumulative reaction profiles, and Figure~\ref{fig:oer_step_deviations} compares the individual reaction-step energies with HSE06/``tight'' results. For uncorrected rSCAN/``tight'', the signed deviations of $\Delta E_{1\text{--}4}^{\mathrm{el}}$ from HSE06/``tight'' are $+0.221$, $-0.177$, $+0.241$, and $-0.362$~eV. ACBN0@rSCAN with the ``tight'' settings reduces these deviations to $+0.151$, $-0.017$, $+0.030$, and $-0.232$~eV, respectively, and lowers their mean absolute deviation from 0.250 to 0.107~eV. The mean absolute deviation from PBE0/``tight'' also decreases, from 0.187~eV for rSCAN to 0.152~eV for ACBN0@rSCAN. With the ``light'' ACBN0@rSCAN settings, the deviations decrease for the first and fourth steps but increase for the second and third steps, giving a mean absolute deviation of 0.225~eV from HSE06/``tight''. Thus, the ``tight'' settings improve agreement for every step, whereas the ``light'' settings give only a modest reduction in the mean absolute deviation, with improvements for two steps and deterioration for the other two. An analogous accuracy assessment is not meaningful for ACBN0@PBE because the PBE-relaxed $\mathrm{*OOH}$ state has a qualitatively different adsorption geometry.

The interpretation of this improvement must account for the additional approximation introduced by the non-variational implementation. The OER energy differences compare clean, adsorbate-covered, and molecular systems with different self-consistent Hubbard parameters, but the derivatives of $\bar U[\rho]$ and $\bar J[\rho]$ are not included in the Kohn--Sham potential. The agreement with the hybrid-functional results is therefore assessed within the present self-consistent ACBN0 procedure.

To examine how the agreement arises, Supporting Information Table~S14 decomposes the deviation of every reaction step from HSE06/``tight'' into slab and molecular-reference contributions for rSCAN and all calculated ACBN0@rSCAN numerical settings. For ACBN0@rSCAN with the ``tight'' settings, the slab and molecular contributions to the second step differ from HSE06/``tight'' by only $+0.018$ and $-0.035$~eV, respectively. In contrast, the first, third, and fourth steps contain larger contributions of opposite sign. For example, the $-1.610$~eV slab deviation in the third step is compensated by a $+1.640$~eV molecular-reference deviation. The smaller net deviations therefore do not demonstrate independent accuracy of the slab and molecular total-energy differences. This cancellation makes the consistent treatment of molecular references essential. Previous DFT+$U$ studies have shown that Hubbard corrections modify the adsorption energies of OER intermediates and the resulting activity trends \cite{xu2015linearresponse,tyminska2017wateroxidation}. Lambert and O'Regan showed for oxygen-vacancy formation that correcting O-$2p$ states in a solid but not in the reference $\mathrm{O}_2$ molecule disrupts cancellation of the Hubbard-energy contributions and can produce unphysical results \cite{lambert2024defects}. Here, the clean slab, adsorbate-covered slabs, and isolated molecules are all treated with the same ACBN0 procedure and numerical settings. The resulting interactions reflect the individual chemical environments, as discussed below.

The mean electronic reaction-step energy, $E_0^{\mathrm{el}}$, provides a complementary measure of the total four-electron reaction energy independently of how it is distributed among the individual steps. HSE06 and PBE0 with the ``tight'' settings give nearly identical values of 1.308 and 1.310~eV, respectively. The rSCAN value is 1.289~eV, and ACBN0@rSCAN with the ``tight'' settings changes it only slightly to 1.291~eV, remaining within $\approx 0.02$~eV of both hybrid-functional results. ACBN0@rSCAN with the ``light'' settings gives 1.306~eV and is also close to the hybrid values. By comparison, PBE and ACBN0@PBE with the ``tight'' settings give 1.250 and 1.252~eV. Thus, ACBN0@rSCAN primarily redistributes the electronic reaction energy among the four steps while producing only a small change in their sum.

The limiting potential $U_L^{\mathrm{el}}$ and electronic-energy-derived overpotential $\eta_{\mathrm{theory}}^{\mathrm{el}}$ in Table~\ref{tab:oer_summary} describe how this reaction energy is distributed, with particular emphasis on the largest uphill step. HSE06 and PBE0 with the ``tight'' settings identify the first step as potential-limiting and give similar values: $U_L^{\mathrm{el}}=2.106$ and 2.100~V and $\eta_{\mathrm{theory}}^{\mathrm{el}}=0.798$ and 0.790~V, respectively. rSCAN gives a larger limiting potential of 2.327~V and an overpotential of 1.038~V. ACBN0@rSCAN with the ``tight'' settings shifts both descriptors toward the hybrid-functional results, reducing them to 2.256 and 0.965~V. The lower values obtained with ACBN0@rSCAN using the ``light'' settings, 1.927 and 0.621~V, should be considered together with its larger step-resolved error because the maximum alone does not measure agreement across the complete reaction profile.

The converged Hubbard parameters in Supporting Information Table~S13 show how the self-consistently evaluated $U$ values change among the clean slab, adsorbate-covered slabs, and molecular references. With the ``tight'' settings, the clean slab has $U_{\mathrm{Ni}\text{-}3d}=4.38$--$4.67$~eV and $U_{\mathrm{O}\text{-}2p}=2.02$--$2.08$~eV. Across the adsorbate-covered slabs, $U_{\mathrm{Ni}\text{-}3d}$ spans $3.55$--$4.43$~eV, while the surface and adsorbate O-$2p$ shells span $1.86$--$2.65$ and $2.21$--$2.65$~eV, respectively. The oxygen-containing molecular references have substantially larger $U_{\mathrm{O}\text{-}2p}$ values of 4.49~eV for $\mathrm{H_2O}$ and 6.13~eV for $\mathrm{O_2}$. These differences highlight the importance of determining the Hubbard parameters self-consistently on an atom-resolved basis, which allows the self-consistent interactions to adapt to chemically inequivalent sites without transferring a single interaction value between distinct environments.

The local magnetic moments provide an additional indication of the basis-set dependence of ACBN0. For ACBN0@rSCAN, the ``light'' numerical settings produce a larger sum of absolute Mulliken moments than the ``tight'' settings for the clean surface and all adsorbate-covered structures, with differences ranging from approximately $0.38$ to $0.96~\mu_B$ per unit cell containing four Ni atoms. This increase is distributed mainly across enhanced local moments on Ni atoms and larger oppositely oriented spin polarization on several surface O atoms; it is not confined to the adsorbate. A control calculation retaining the ``light'' basis while using the ``tight'' integration grids changes the sums of absolute moments by only $0.004$--$0.046~\mu_B$ and the individual OER reaction-step energies by $0.005$--$0.032$~eV, with a mean absolute change of $0.019$~eV. The control retains a ``light''-basis-like reaction-step error profile, confirming that grid refinement alone does not account for the larger differences between the complete ``light'' and ``tight'' settings. The stronger local spin polarization obtained with the ``light'' basis is consistent with excessive localization by the Hubbard correction. The additional radial flexibility of the ``tight'' basis reduces this effect and brings the individual OER reaction-step energies closer to the hybrid-functional results.

\section{Conclusions}\label{sec:conclusion}
We formulated and implemented the ACBN0 pseudo-hybrid density functional in a numeric atom-centered orbital basis within the all-electron, full-potential electronic-structure package FHI-aims. A L\"{o}wdin-orthogonalized projector based on the ``minimal'' basis set functions provides stable occupation and renormalized density matrices for the tested systems, while the interactions $\bar U$ and $\bar J$ are updated within the SCF cycle and can be combined with the FLL, AMF, and Petukhov double-counting treatments. An individual ACBN0 iteration remains close in cost to a semi-local iteration for the tested structures. The total ACBN0 cost, however, depends on the additional iterations required to converge the coupled density and interaction update.

Across the diagnostic set of bulk oxides and nitrides, FLL gives the most consistent combined description of band gaps and magnetic moments. With the midpoint of the reported experimental span used as the reference for both quantities, ACBN0@SCAN--FLL yields the lowest band gap MAPE among the tested ACBN0 combinations, 20.5\%, comparable to the HSE06 reference. ACBN0 reduces the experimental MAPE of the magnetic moments from 37.0\% for PBE and 33.0--33.7\% for SCAN to 13.9--16.9\%. The preferred parent functional nevertheless remains material-dependent: stronger localization with SCAN can improve early $3d$ oxides but can overlocalize late transition metal oxides with nearly filled shells.

For adsorption and OER on $\beta$-NiOOH(001), atom-resolved ACBN0 assigns separate interactions to chemically inequivalent surface, adsorbate, and molecular-reference atoms. With the ``tight'' settings, ACBN0@rSCAN reduces the mean absolute deviation from HSE06 of the four electronic reaction-step energies from 0.250 to 0.107~eV. The decomposition of these differences also reveals substantial cancellation between slab and molecular-reference contributions, so this improvement applies to the consistently calculated reaction profile and does not establish independent accuracy of its components. Consistent treatment of the surface, adsorbates, and molecular references is therefore essential. The closer agreement between ACBN0 and HSE06 found here concerns the electronic reaction energies. The associated overpotential estimates use a method-specific electronic reference and do not include the thermodynamic contributions needed for a free-energy-based assessment. Zero-point-energy and entropy contributions may affect both the absolute overpotentials and the quantitative differences between the methods. These contributions were not evaluated, as the present benchmark is restricted to single-point electronic energies.

The principal practical limitation is the coupled dependence on the localized projector, basis set, parent functional, and target property. The ``light'' basis produces stronger localization and smaller aggregated band gap errors in the bulk benchmark, whereas the ``tight'' basis gives closer agreement with the hybrid-functional surface reaction profiles. Replacing only the integration grids does not recover the ``tight''-basis result. Robust construction of the Hubbard subspace remains an important task for the ACBN0 procedure.

The present work establishes a practical formulation of ACBN0 in a localized-orbital basis and provides a foundation for its further development. A fully variational formulation that includes the density dependence of the Hubbard parameters is also needed to place total energies and energy differences on a consistent variational footing. Implementing analytical forces and stresses would enable fully self-consistent structural relaxation, while additional mixing schemes for the Hubbard parameters could accelerate and stabilize the coupled SCF--ACBN0 cycle. Extending the implementation to alternative localized subspaces, including Wannier functions and natural atomic orbitals, could further improve its numerical robustness by reducing the sensitivity of the results to the basis set.

\bibliographystyle{elsarticle-num}
\bibliography{apssamp}

\clearpage
\onecolumn
\section*{Supporting Information}
\setcounter{table}{0}
\setcounter{figure}{0}
\setcounter{equation}{0}
\renewcommand{\theequation}{S\arabic{equation}}
\renewcommand{\theHtable}{S\arabic{table}}
\renewcommand{\theHfigure}{S\arabic{figure}}
\renewcommand{\theHequation}{S\arabic{equation}}

\renewcommand{\thetable}{S\arabic{table}}
\renewcommand{\thefigure}{S\arabic{figure}}

\begin{figure}[htbp]
\centering
\includegraphics[width=0.92\textwidth]{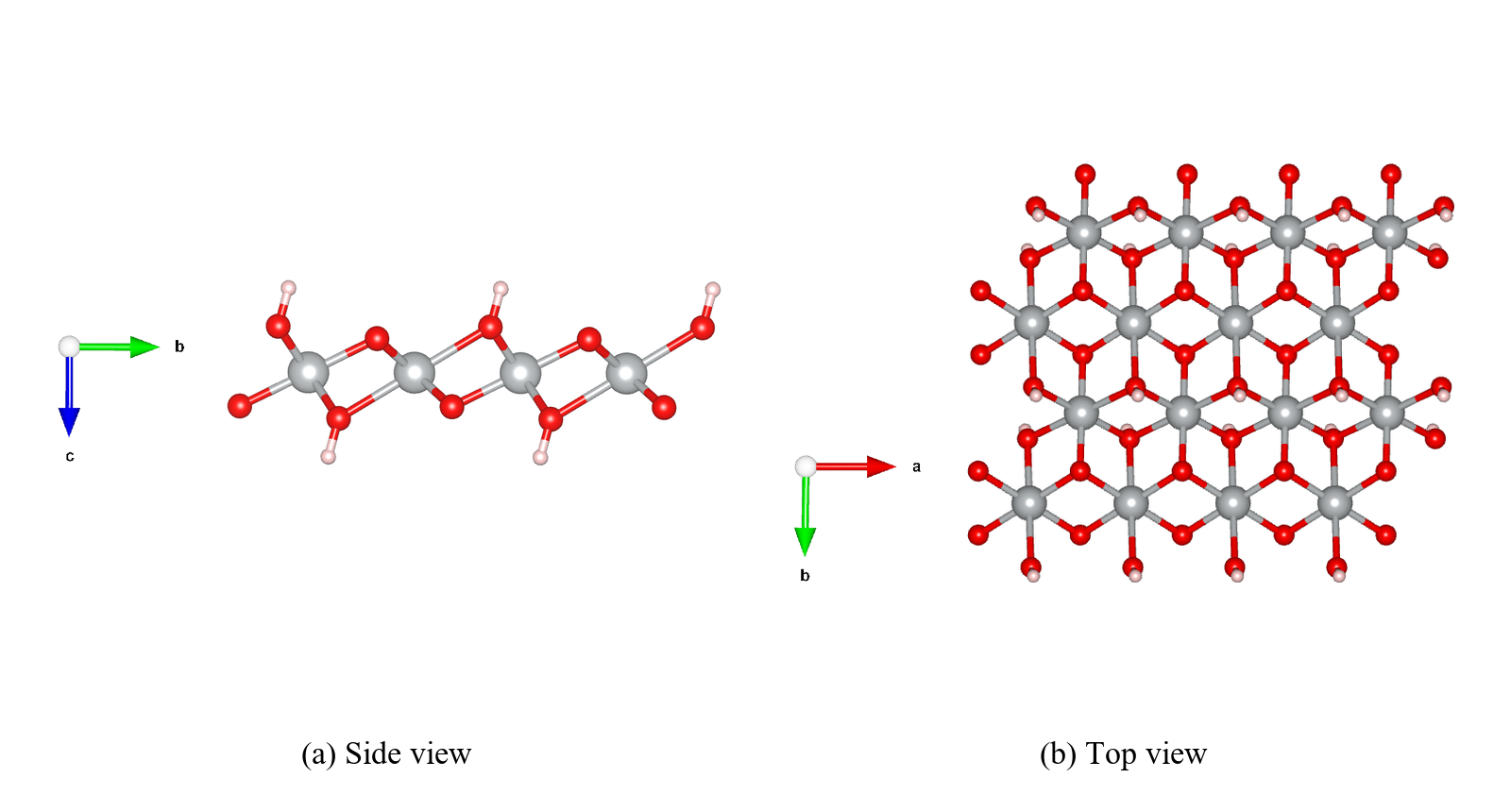}
\caption{Side and top views of the stoichiometric monolayer slab model of the $\beta$-NiOOH(001) surface. The unit cell contains four NiOOH formula units (periodic replicas are shown). The HSE06-relaxed in-plane lattice-vector lengths are $a=5.775$~\AA\ and $b=5.115$~\AA, with an angle of $\gamma=90.8^{\circ}$ between them. The out-of-plane lattice-vector length is 50.0~\AA\ and includes the vacuum region. Gray, red, and white spheres denote Ni, O, and H atoms, respectively.}
\label{si:fig:niooh_surface_cell}
\end{figure}

\clearpage

\begin{table}[htbp]
\centering
\caption{Lattice parameters calculated with PBE using the ``tight'' numerical settings and the listed $\bm{k}$-point meshes.}
\label{si:tab:structure}
\normalsize
\begin{tabular}{lccccccc}
\toprule
\textbf{Material} & \textbf{$k$-grid} & \multicolumn{3}{c}{\textbf{Lattice parameters (\AA)}} & \multicolumn{3}{c}{\textbf{Angles ($^\circ$)}} \\
\cmidrule(lr){3-5} \cmidrule(lr){6-8}
 & & \textbf{$a$} & \textbf{$b$} & \textbf{$c$} & \textbf{$\alpha$} & \textbf{$\beta$} & \textbf{$\gamma$} \\
\midrule
\multicolumn{8}{l}{\textbf{\textit{3d transition metal oxides}}} \\
\midrule
CoO & $8\times8\times4$ & 2.830 & 2.818 & 5.585 & 104.54 & 90.00 & 120.14 \\
Cr$_2$O$_3$ & $6\times6\times6$ & 4.939 & 5.412 & 4.939 & 117.15 & 120.00 & 90.00 \\
Cu$_2$O & $7\times7\times7$ & 4.299 & 4.299 & 4.299 & 90.00 & 90.00 & 90.00 \\
CuO & $11\times11\times6$ & 2.944 & 2.944 & 5.158 & 90.00 & 90.00 & 90.00 \\
MnO & $6\times6\times4$ & 3.181 & 3.181 & 5.291 & 90.00 & 72.49 & 60.01 \\
NiO & $8\times8\times6$ & 2.970 & 2.971 & 5.115 & 90.02 & 106.88 & 119.99 \\
TiO$_2$ (rut.) & $6\times6\times5$ & 4.641 & 4.641 & 2.964 & 90.00 & 90.00 & 90.00 \\
TiO$_2$ (an.) & $8\times8\times3$ & 3.798 & 3.798 & 9.704 & 90.00 & 90.00 & 90.00 \\
\midrule
\multicolumn{8}{l}{\textbf{\textit{Alkaline earth and post-transition metal oxides}}} \\
\midrule
BaO & $6\times6\times6$ & 5.569 & 5.569 & 5.569 & 90.00 & 90.00 & 90.00 \\
BeO & $12\times12\times7$ & 2.712 & 2.712 & 4.404 & 90.00 & 90.00 & 120.00 \\
$\alpha$-Bi$_2$O$_3$ & $5\times4\times4$ & 5.932 & 8.324 & 7.555 & 90.00 & 112.59 & 90.00 \\
CaO & $7\times7\times7$ & 4.833 & 4.833 & 4.833 & 90.00 & 90.00 & 90.00 \\
CdO & $7\times7\times7$ & 4.765 & 4.765 & 4.765 & 90.00 & 90.00 & 90.00 \\
HgO & $5\times6\times9$ & 6.687 & 5.851 & 3.740 & 90.00 & 90.00 & 90.00 \\
MgO & $8\times8\times8$ & 4.255 & 4.255 & 4.255 & 90.00 & 90.00 & 90.00 \\
SrO & $6\times6\times6$ & 5.194 & 5.194 & 5.194 & 90.00 & 90.00 & 90.00 \\
ZnO & $8\times8\times6$ & 3.278 & 3.278 & 5.289 & 90.00 & 90.00 & 120.00 \\
\midrule
\multicolumn{8}{l}{\textbf{\textit{Nitrides}}} \\
\midrule
AlN & $10\times10\times6$ & 3.131 & 3.131 & 5.017 & 90.00 & 90.00 & 120.00 \\
GaN & $10\times10\times6$ & 3.219 & 3.219 & 5.242 & 90.00 & 90.00 & 120.00 \\
\bottomrule
\end{tabular}
\end{table}

\begin{table}[htbp]
  \centering
  \caption{Experimental and calculated band gaps for the complete test set obtained with HSE06, PBE, and ACBN0@PBE (eV). Materials are grouped by chemical family. AFM denotes an antiferromagnetic configuration. ``l'' and ``t'' denote the ``light'' and ``tight'' numerical settings. ``rel'' denotes HSE06 results on HSE06-relaxed geometries; all other values are from single-point calculations on geometries relaxed with PBE using the ``tight'' settings. All ACBN0 calculations use L\"{o}wdin projectors.}
  \label{si:tab:SI_PBE_methods}
  \normalsize
  \setlength{\tabcolsep}{2pt}
  \renewcommand{\arraystretch}{1.05}
  \begin{tabular}{lcccccccccccc}
    \toprule
    & & \multicolumn{3}{c}{\textbf{HSE06}} & \multicolumn{2}{c}{\textbf{PBE}} & \multicolumn{6}{c}{\textbf{ACBN0@PBE}} \\
    & & & & & & & \multicolumn{2}{c}{Petukhov} & \multicolumn{2}{c}{FLL} & \multicolumn{2}{c}{AMF} \\
    \cmidrule(lr){3-5} \cmidrule(lr){6-7} \cmidrule(lr){8-9} \cmidrule(lr){10-11} \cmidrule(lr){12-13}
    \textbf{Material} & \textbf{Exp.}  & \makebox[2em]{rel} & \makebox[2em]{l} & \makebox[2em]{t} & \makebox[2em]{l} & \makebox[2em]{t} & \makebox[2em]{l} & \makebox[2em]{t} & \makebox[2em]{l} & \makebox[2em]{t} & \makebox[2em]{l} & \makebox[2em]{t} \\
    \midrule
    \multicolumn{13}{l}{\textbf{\textit{3d transition metal oxides}}} \\
    \midrule
    \ce{TiO2} (rut.)    & 3.05 \cite{Cronemeyer1952} , 3.061-3.065 \cite{Pascual1978}& \makebox[2em]{3.27} & \makebox[2em]{3.18} & \makebox[2em]{3.29} & \makebox[2em]{1.76} & \makebox[2em]{1.84} & \makebox[2em]{1.93} & \makebox[2em]{1.92} & \makebox[2em]{2.28} & \makebox[2em]{2.02} & \makebox[2em]{1.93} & \makebox[2em]{1.92} \\
    \ce{TiO2} (an.)    & 3.3 \cite{tang1993photoluminescence}, 3.3-3.4 \cite{reddy2003bandgap} & \makebox[2em]{3.62} & \makebox[2em]{3.50} & \makebox[2em]{3.58} & \makebox[2em]{2.09} & \makebox[2em]{2.13} & \makebox[2em]{2.24} & \makebox[2em]{2.21} & \makebox[2em]{2.56} & \makebox[2em]{2.35} & \makebox[2em]{2.23} & \makebox[2em]{2.21} \\
    \ce{Cr2O3} (AFM)    & 3.2 \cite{10.1063/1.4867012}, 3.0 \cite{SINGH2019282} & \makebox[2em]{4.54} & \makebox[2em]{4.46} & \makebox[2em]{4.49} & \makebox[2em]{1.61} & \makebox[2em]{1.62} & \makebox[2em]{2.70} & \makebox[2em]{2.45} & \makebox[2em]{2.72} & \makebox[2em]{2.51} & \makebox[2em]{2.48} & \makebox[2em]{2.31} \\
    \ce{MnO} (AFM)      & 3.9 \cite{PhysRevB.44.1530}, 4.1 \cite{kurmaev2008oxygen}  & \makebox[2em]{2.86} & \makebox[2em]{2.85} & \makebox[2em]{2.89} & \makebox[2em]{0.83} & \makebox[2em]{0.85} & \makebox[2em]{0.80} & \makebox[2em]{0.79} & \makebox[2em]{2.01} & \makebox[2em]{1.70} & \makebox[2em]{0.71} & \makebox[2em]{0.76} \\
    \ce{CoO} (AFM)      & 2.6 \cite{kurmaev2008oxygen}, 2.8 \cite{pratt1959optical} & \makebox[2em]{3.61} & \makebox[2em]{2.56} & \makebox[2em]{2.49} & \makebox[2em]{0.00} & \makebox[2em]{0.00} & \makebox[2em]{2.28} & \makebox[2em]{1.83} & \makebox[2em]{2.60} & \makebox[2em]{1.94} & \makebox[2em]{1.76} & \makebox[2em]{1.55} \\
    \ce{NiO} (AFM)      &  4.0 \cite{kurmaev2008oxygen,PhysRevB.2.2182}, 4.3 \cite{PhysRevLett.53.2339} & \makebox[2em]{4.41} & \makebox[2em]{4.38} & \makebox[2em]{4.40} & \makebox[2em]{0.99} & \makebox[2em]{0.97} & \makebox[2em]{3.08} & \makebox[2em]{2.76} & \makebox[2em]{3.67} & \makebox[2em]{3.07} & \makebox[2em]{2.41} & \makebox[2em]{2.40} \\
    \ce{Cu2O}           & 2.4 \cite{PhysRevB.38.11322} & \makebox[2em]{1.98} & \makebox[2em]{1.97} & \makebox[2em]{1.96} & \makebox[2em]{0.48} & \makebox[2em]{0.48} & \makebox[2em]{0.76} & \makebox[2em]{0.80} & \makebox[2em]{1.42} & \makebox[2em]{1.08} & \makebox[2em]{0.68} & \makebox[2em]{0.70} \\
    \ce{CuO} (AFM)      & 1.35 \cite{koffyberg1982photoelectrochemical}, 1.40  \cite{PhysRevB.38.11322}&  \makebox[2em]{1.97} & \makebox[2em]{1.97} & \makebox[2em]{2.01} & \makebox[2em]{0.00} & \makebox[2em]{0.00} & \makebox[2em]{1.33} & \makebox[2em]{1.23} & \makebox[2em]{1.92} & \makebox[2em]{1.67} & \makebox[2em]{1.08} & \makebox[2em]{1.07} \\
    \midrule
    \multicolumn{13}{l}{\textbf{\textit{Alkaline earth oxides and post-transition metal oxides}}} \\
    \midrule
    \ce{BeO}         & 10.63 \cite{ROESSLER1969157} &  \makebox[2em]{9.62} & \makebox[2em]{9.40} & \makebox[2em]{9.33} & \makebox[2em]{7.52} & \makebox[2em]{7.46} & \makebox[2em]{7.52} & \makebox[2em]{7.46} & \makebox[2em]{7.95} & \makebox[2em]{7.47} & \makebox[2em]{7.52} & \makebox[2em]{7.46} \\
    \ce{MgO}         &  7.77 \cite{PhysRev.159.733}, 7.833 \cite{PhysRevLett.22.1428}, 6.4 \cite{PhysRevB.81.245123} & \makebox[2em]{6.36} & \makebox[2em]{6.10} & \makebox[2em]{6.17} & \makebox[2em]{4.38} & \makebox[2em]{4.44} & \makebox[2em]{4.38} & \makebox[2em]{4.44} & \makebox[2em]{5.80} & \makebox[2em]{4.71} & \makebox[2em]{4.38} & \makebox[2em]{4.44} \\
    \ce{CaO}         &  7.034 \cite{PhysRevLett.22.1428}, 6.875 \cite{kaneko1988new}, 7.085 \cite{PhysRevLett.22.1428} & \makebox[2em]{5.14} & \makebox[2em]{5.14} & \makebox[2em]{5.25} & \makebox[2em]{3.54} & \makebox[2em]{3.63} & \makebox[2em]{3.54} & \makebox[2em]{3.63} & \makebox[2em]{4.86} & \makebox[2em]{3.99} & \makebox[2em]{3.54} & \makebox[2em]{3.63} \\
    \ce{SrO}         &  5.77 \cite{PhysRev.113.1019}, 5.793 \cite{kaneko1988new} & \makebox[2em]{4.76} & \makebox[2em]{4.77} & \makebox[2em]{4.75} & \makebox[2em]{3.31} & \makebox[2em]{3.30} & \makebox[2em]{3.31} & \makebox[2em]{3.30} & \makebox[2em]{3.97} & \makebox[2em]{3.67} & \makebox[2em]{3.31} & \makebox[2em]{3.30} \\
    \ce{BaO}         &  4.1 \cite{PhysRev.113.1019}, 3.985 \cite{kaneko1988new}, 5.1 \cite{vratny1962reflectance}  & \makebox[2em]{3.21} & \makebox[2em]{3.23} & \makebox[2em]{3.27} & \makebox[2em]{2.01} & \makebox[2em]{2.05} & \makebox[2em]{2.01} & \makebox[2em]{2.05} & \makebox[2em]{2.99} & \makebox[2em]{2.54} & \makebox[2em]{2.01} & \makebox[2em]{2.05} \\
    \ce{ZnO}         &  3.3 \cite{10.1063/1.367375}, 3.345 \cite{PhysRev.143.512} & \makebox[2em]{2.42} & \makebox[2em]{2.33} & \makebox[2em]{2.33} & \makebox[2em]{0.75} & \makebox[2em]{0.75} & \makebox[2em]{0.75} & \makebox[2em]{0.74} & \makebox[2em]{1.98} & \makebox[2em]{1.77} & \makebox[2em]{0.74} & \makebox[2em]{0.74} \\
    \ce{CdO}         &  1.2 \cite{mcguinness2003influence}, 0.84 \cite{koffyberg1976thermoreflectance} & \makebox[2em]{0.86} & \makebox[2em]{0.77} & \makebox[2em]{0.78} & \makebox[2em]{0.43} & \makebox[2em]{0.43} & \makebox[2em]{0.42} & \makebox[2em]{0.42} & \makebox[2em]{0.44} & \makebox[2em]{0.19} & \makebox[2em]{0.42} & \makebox[2em]{0.42} \\
    \ce{HgO}         &  1.9 \cite{DASILVAPEREIRA1982167,PhysRevB.71.235109}, $>$2.13 \cite{PhysRevB.57.153} & \makebox[2em]{2.43} & \makebox[2em]{2.54} & \makebox[2em]{2.56} & \makebox[2em]{1.34} & \makebox[2em]{1.36} & \makebox[2em]{1.61} & \makebox[2em]{1.56} & \makebox[2em]{1.76} & \makebox[2em]{1.52} & \makebox[2em]{1.60} & \makebox[2em]{1.56} \\
    \ce{$\alpha$-Bi2O3}       &  2.80 \cite{cheng2010synergistic}, 2.734 \cite{IYYAPUSHPAM2013104} &  \makebox[2em]{3.20} & \makebox[2em]{3.22} & \makebox[2em]{3.22} & \makebox[2em]{2.27} & \makebox[2em]{2.27} & \makebox[2em]{2.32} & \makebox[2em]{2.30} & \makebox[2em]{2.73} & \makebox[2em]{2.55} & \makebox[2em]{2.32} & \makebox[2em]{2.30} \\
    \midrule
    \multicolumn{13}{l}{\textbf{\textit{Nitrides}}} \\
    \midrule
    \ce{GaN}    & 3.474 \cite{PhysRevB.4.1211}, 3.6 \cite{bloom1974band}, 3.44 \cite{Perlin_1993} &  \makebox[2em]{3.17} & \makebox[2em]{2.95} & \makebox[2em]{2.92} & \makebox[2em]{1.76} & \makebox[2em]{1.72} & \makebox[2em]{1.76} & \makebox[2em]{1.72} & \makebox[2em]{1.92} & \makebox[2em]{1.85} & \makebox[2em]{1.76} & \makebox[2em]{1.72} \\
    \ce{AlN}     & 6.2 \cite{yim1973epitaxially}, 6.28 \cite{perry1978optical},  6.12 \cite{li2003band} & \makebox[2em]{5.65} & \makebox[2em]{5.49} & \makebox[2em]{5.43} & \makebox[2em]{4.09} & \makebox[2em]{4.03} & \makebox[2em]{4.09} & \makebox[2em]{4.03} & \makebox[2em]{4.06} & \makebox[2em]{3.97} & \makebox[2em]{4.09} & \makebox[2em]{4.03} \\

    \bottomrule
  \end{tabular}
\end{table}

\begin{table}[htbp]
  \centering
  \caption{Calculated band gaps for the complete test set obtained with SCAN and ACBN0@SCAN (eV). Materials are grouped by chemical family. AFM denotes an antiferromagnetic configuration. ``l'' and ``t'' denote the ``light'' and ``tight'' numerical settings. All values are from single-point calculations on geometries relaxed with PBE using the ``tight'' settings. All ACBN0 calculations use L\"{o}wdin projectors.}
  \label{si:tab:SI_SCAN_methods}
  \normalsize
  \setlength{\tabcolsep}{3pt}
  \renewcommand{\arraystretch}{1.05}
  \begin{tabular}{l cc cc cc cc}
    \toprule
     & \multicolumn{2}{c}{\textbf{SCAN}} & \multicolumn{6}{c}{\textbf{ACBN0@SCAN}} \\ 
     & &  &  \multicolumn{2}{c}{Petukhov} & \multicolumn{2}{c}{FLL} & \multicolumn{2}{c}{AMF} \\
    \cmidrule(lr){2-3} \cmidrule(lr){4-5} \cmidrule(lr){6-7} \cmidrule(lr){8-9}
    \textbf{Material}  & \makebox[2.3em]{l} & \makebox[2.3em]{t} & \makebox[2.3em]{l} & \makebox[2.3em]{t} & \makebox[2.3em]{l} & \makebox[2.3em]{t} & \makebox[2.3em]{l} & \makebox[2.3em]{t} \\
    \midrule
    \multicolumn{9}{l}{\textbf{\textit{3d transition metal oxides}}} \\
    \midrule
    \ce{TiO2} (rut.)    & \makebox[2.3em]{2.11} & \makebox[2.3em]{2.19} & \makebox[2.3em]{2.31} & \makebox[2.3em]{2.28} & \makebox[2.3em]{2.71} & \makebox[2.3em]{2.39} & \makebox[2.3em]{2.31} & \makebox[2.3em]{2.28} \\
    \ce{TiO2} (an.)    & \makebox[2.3em]{2.49} & \makebox[2.3em]{2.53} & \makebox[2.3em]{2.66} & \makebox[2.3em]{2.62} & \makebox[2.3em]{3.03} & \makebox[2.3em]{2.77} & \makebox[2.3em]{2.66} & \makebox[2.3em]{2.62} \\
    \ce{Cr2O3} (AFM)    & \makebox[2.3em]{2.97} & \makebox[2.3em]{3.00} & \makebox[2.3em]{3.79} & \makebox[2.3em]{3.65} & \makebox[2.3em]{3.88} & \makebox[2.3em]{3.76} & \makebox[2.3em]{3.59} & \makebox[2.3em]{3.51} \\
    \ce{MnO} (AFM)      & \makebox[2.3em]{1.54} & \makebox[2.3em]{1.58} & \makebox[2.3em]{1.53} & \makebox[2.3em]{1.54} & \makebox[2.3em]{2.70} & \makebox[2.3em]{2.38} & \makebox[2.3em]{1.48} & \makebox[2.3em]{1.52} \\
    \ce{CoO} (AFM)      & \makebox[2.3em]{0.00} & \makebox[2.3em]{0.27} & \makebox[2.3em]{2.99} & \makebox[2.3em]{2.48} & \makebox[2.3em]{3.21} & \makebox[2.3em]{2.64} & \makebox[2.3em]{2.42} & \makebox[2.3em]{2.23} \\
    \ce{NiO} (AFM)      & \makebox[2.3em]{2.60} & \makebox[2.3em]{2.74} & \makebox[2.3em]{3.76} & \makebox[2.3em]{3.45} & \makebox[2.3em]{4.41} & \makebox[2.3em]{3.76} & \makebox[2.3em]{3.11} & \makebox[2.3em]{3.07} \\
    \ce{Cu2O}           & \makebox[2.3em]{0.77} & \makebox[2.3em]{0.75} & \makebox[2.3em]{1.00} & \makebox[2.3em]{1.03} & \makebox[2.3em]{1.69} & \makebox[2.3em]{1.33} & \makebox[2.3em]{0.76} & \makebox[2.3em]{0.74} \\
    \ce{CuO} (AFM)      & \makebox[2.3em]{0.00} & \makebox[2.3em]{0.00} & \makebox[2.3em]{1.70} & \makebox[2.3em]{1.61} & \makebox[2.3em]{2.19} & \makebox[2.3em]{1.94} & \makebox[2.3em]{1.48} & \makebox[2.3em]{1.48} \\
    \midrule
    \multicolumn{9}{l}{\textbf{\textit{Alkaline earth oxides and post-transition metal oxides}}} \\
    \midrule
    \ce{BeO}         & \makebox[2.3em]{8.39} & \makebox[2.3em]{8.29} & \makebox[2.3em]{8.39} & \makebox[2.3em]{8.29} & \makebox[2.3em]{8.87} & \makebox[2.3em]{8.32} & \makebox[2.3em]{8.39} & \makebox[2.3em]{8.29} \\
    \ce{MgO}          & \makebox[2.3em]{5.26} & \makebox[2.3em]{5.31} & \makebox[2.3em]{5.26} & \makebox[2.3em]{5.31} & \makebox[2.3em]{6.82} & \makebox[2.3em]{5.61} & \makebox[2.3em]{5.26} & \makebox[2.3em]{5.31} \\
    \ce{CaO}          & \makebox[2.3em]{4.11} & \makebox[2.3em]{4.17} & \makebox[2.3em]{4.11} & \makebox[2.3em]{4.17} & \makebox[2.3em]{5.50} & \makebox[2.3em]{4.55} & \makebox[2.3em]{4.11} & \makebox[2.3em]{4.17} \\
    \ce{SrO}          & \makebox[2.3em]{3.79} & \makebox[2.3em]{3.77} & \makebox[2.3em]{3.79} & \makebox[2.3em]{3.77} & \makebox[2.3em]{4.48} & \makebox[2.3em]{4.16} & \makebox[2.3em]{3.79} & \makebox[2.3em]{3.77} \\
    \ce{BaO}          & \makebox[2.3em]{2.35} & \makebox[2.3em]{2.38} & \makebox[2.3em]{2.35} & \makebox[2.3em]{2.38} & \makebox[2.3em]{3.41} & \makebox[2.3em]{2.90} & \makebox[2.3em]{2.35} & \makebox[2.3em]{2.38} \\
    \ce{ZnO}          & \makebox[2.3em]{1.12} & \makebox[2.3em]{1.11} & \makebox[2.3em]{1.12} & \makebox[2.3em]{1.10} & \makebox[2.3em]{2.40} & \makebox[2.3em]{2.17} & \makebox[2.3em]{1.12} & \makebox[2.3em]{1.10} \\
    \ce{CdO}         & \makebox[2.3em]{0.76} & \makebox[2.3em]{0.76} & \makebox[2.3em]{0.75} & \makebox[2.3em]{0.75} & \makebox[2.3em]{0.91} & \makebox[2.3em]{0.64} & \makebox[2.3em]{0.75} & \makebox[2.3em]{0.74} \\
    \ce{HgO}         & \makebox[2.3em]{1.56} & \makebox[2.3em]{1.61} & \makebox[2.3em]{1.84} & \makebox[2.3em]{1.82} & \makebox[2.3em]{1.97} & \makebox[2.3em]{1.74} & \makebox[2.3em]{1.84} & \makebox[2.3em]{1.82} \\
    \ce{$\alpha$-Bi2O3}     & \makebox[2.3em]{2.47} & \makebox[2.3em]{2.49} & \makebox[2.3em]{2.51} & \makebox[2.3em]{2.51} & \makebox[2.3em]{2.95} & \makebox[2.3em]{2.77} & \makebox[2.3em]{2.51} & \makebox[2.3em]{2.51} \\
    \midrule
    \multicolumn{9}{l}{\textbf{\textit{Nitrides}}} \\
    \midrule
    \ce{GaN}  & \makebox[2.3em]{2.06} & \makebox[2.3em]{2.04} & \makebox[2.3em]{2.08} & \makebox[2.3em]{2.04} & \makebox[2.3em]{2.23} & \makebox[2.3em]{2.19} & \makebox[2.3em]{2.08} & \makebox[2.3em]{2.04} \\
    \ce{AlN}      & \makebox[2.3em]{4.76} & \makebox[2.3em]{4.67} & \makebox[2.3em]{4.75} & \makebox[2.3em]{4.67} & \makebox[2.3em]{4.73} & \makebox[2.3em]{4.61} & \makebox[2.3em]{4.75} & \makebox[2.3em]{4.67} \\
    \bottomrule
  \end{tabular}
\end{table}

\begin{table}[htbp]
  \centering
  \caption{Experimental and calculated Mulliken magnetic moments per metal atom for antiferromagnetic (AFM) transition metal oxides obtained with HSE06, PBE, and ACBN0@PBE ($\mu_B$). ``l'' and ``t'' denote the ``light'' and ``tight'' numerical settings. ``rel'' denotes HSE06 results on HSE06-relaxed geometries; all other values are from single-point calculations on geometries relaxed with PBE using the ``tight'' settings. All ACBN0 calculations use L\"{o}wdin projectors. The aggregate experimental errors in the main text use the midpoint of the full span of the listed experimental values. A single reported value is used unchanged.}
  \label{si:tab:magn_mom_PBE_methods}
  \normalsize
  \setlength{\tabcolsep}{2pt}
  \renewcommand{\arraystretch}{1.05}
  \begin{tabular}{lcccccccccccc}
    \toprule
    & & \multicolumn{3}{c}{\textbf{HSE06}} & \multicolumn{2}{c}{\textbf{PBE}} & \multicolumn{6}{c}{\textbf{ACBN0@PBE}} \\
    & & & & & & & \multicolumn{2}{c}{Petukhov} & \multicolumn{2}{c}{FLL} & \multicolumn{2}{c}{AMF} \\
    \cmidrule(lr){3-5} \cmidrule(lr){6-7} \cmidrule(lr){8-9} \cmidrule(lr){10-11} \cmidrule(lr){12-13}
    \textbf{Material} & \textbf{Exp.}  & \makebox[2em]{rel} & \makebox[2em]{l} & \makebox[2em]{t} & \makebox[2em]{l} & \makebox[2em]{t} & \makebox[2em]{l} & \makebox[2em]{t} & \makebox[2em]{l} & \makebox[2em]{t} & \makebox[2em]{l} & \makebox[2em]{t} \\
    \midrule
    \ce{Cr2O3} & 2.48 \cite{brown2002determination}, 2.76 \cite{skovhus2022magnons} & 3.04 & 3.03 & 3.01 & 2.88 & 2.87 & 2.89 & 2.87 & 3.14 & 3.10 & 2.78 & 2.80 \\
    \ce{MnO}   & 4.89 \cite{BONFANTE1972553}, 5.0 \cite{PhysRev.110.1333} & 4.80 & 4.81 & 4.84 & 4.63 & 4.69 & 4.61 & 4.68 & 4.96 & 5.04 & 4.58 & 4.66 \\
    \ce{CoO}   & 3.8 \cite{PhysRev.110.1333}, 3.98 \cite{PhysRevB.64.052102} & 2.72 & 2.68 & 2.63 & 2.39 & 2.35 & 2.76 & 2.70 & 2.80 & 2.75 & 2.70 & 2.63 \\
    \ce{NiO}   & 1.9 \cite{PhysRevB.27.6964}, 2.0 \cite{PhysRev.110.1333} & 1.67 & 1.67 & 1.66 & 1.36 & 1.35 & 1.92 & 1.80 & 1.93 & 1.84 & 1.80 & 1.75 \\
    \ce{CuO}   & 0.69 \cite{PhysRevB.39.4343} & 0.63 & 0.63 & 0.63 & 0.00 & 0.00 & 0.93 & 0.90 & 0.83 & 0.82 & 0.90 & 0.85 \\
    \bottomrule
  \end{tabular}
\end{table}

\begin{table}[htbp]
  \centering
  \caption{Calculated Mulliken magnetic moments per metal atom for antiferromagnetic (AFM) transition metal oxides obtained with SCAN and ACBN0@SCAN ($\mu_B$). ``l'' and ``t'' denote the ``light'' and ``tight'' numerical settings. All values are from single-point calculations on geometries relaxed with PBE using the ``tight'' settings. All ACBN0 calculations use L\"{o}wdin projectors. The two values reported for CoO with ACBN0@SCAN--FLL are the magnitudes of the moments on the two crystallographically equivalent Co sites; their mean is used in the aggregate error analysis.}
  \label{si:tab:magn_mom_SCAN_methods}
  \normalsize
  \setlength{\tabcolsep}{2pt}
  \renewcommand{\arraystretch}{1.05}
  \begin{tabular}{lcccccccc}
    \toprule
    & \multicolumn{2}{c}{\textbf{SCAN}} & \multicolumn{6}{c}{\textbf{ACBN0@SCAN}} \\
    & & & \multicolumn{2}{c}{Petukhov} & \multicolumn{2}{c}{FLL} & \multicolumn{2}{c}{AMF} \\
    \cmidrule(lr){2-3} \cmidrule(lr){4-5} \cmidrule(lr){6-7} \cmidrule(lr){8-9}
    \textbf{Material} & \makebox[2em]{l} & \makebox[2em]{t} & \makebox[2em]{l} & \makebox[2em]{t} & \makebox[2em]{l} & \makebox[2em]{t} & \makebox[2em]{l} & \makebox[2em]{t} \\
    \midrule
    \ce{Cr2O3} & 2.94 & 2.88 & 2.91 & 2.85 & 3.06 & 2.99 & 2.83 & 2.80 \\
    \ce{MnO}   & 4.75 & 4.76 & 4.74 & 4.76 & 4.99 & 5.02 & 4.72 & 4.75 \\
    \ce{CoO}   & 2.54 & 2.54 & 2.85 & 2.76 & 2.85/2.93 & 2.82/2.83 & 2.78 & 2.70 \\
    \ce{NiO}   & 1.61 & 1.60 & 1.96 & 1.88 & 1.96 & 1.95 & 1.86 & 1.83 \\
    \ce{CuO}   & 0.00 & 0.01 & 0.97 & 0.96 & 0.94 & 0.93 & 0.95 & 0.91 \\
    \bottomrule
  \end{tabular}
\end{table}

\begin{table}[htbp]
  \centering
  \caption{Hubbard $U$ values for the listed atomic shells (eV) from ACBN0@PBE with the L\"{o}wdin projector. ``l'' and ``t'' denote the ``light'' and ``tight'' numerical settings. All calculations are single-point calculations on geometries relaxed with the PBE ``tight'' settings.}
  \label{si:tab:SI_U_Me_PBE}
  \normalsize
  \setlength{\tabcolsep}{3pt}
  \renewcommand{\arraystretch}{1.05}
  \begin{tabular}{l c cc cc cc}
    \toprule
    & & \multicolumn{6}{c}{\textbf{ACBN0@PBE}} \\
    & & \multicolumn{2}{c}{Petukhov} & \multicolumn{2}{c}{FLL} & \multicolumn{2}{c}{AMF} \\
    \cmidrule(lr){3-4} \cmidrule(lr){5-6} \cmidrule(lr){7-8}
    \textbf{Material} & \textbf{Shell}& \makebox[2.3em]{l} & \makebox[2.3em]{t} & \makebox[2.3em]{l} & \makebox[2.3em]{t} & \makebox[2.3em]{l} & \makebox[2.3em]{t} \\
    \midrule
    \multicolumn{8}{l}{\textbf{\textit{3d transition metal oxides}}} \\
    \midrule
    \ce{TiO2} (rut.) & 3d  & 0.49 & 0.38 & 0.45 & 0.36 & 0.49 & 0.38 \\
    \ce{TiO2} (an.) & 3d & 0.51 & 0.39 & 0.46 & 0.37 & 0.51 & 0.39 \\
    \ce{Cr2O3} (AFM) & 3d & 2.81 & 2.55 & 2.66 & 2.44 & 2.91 & 2.60 \\
    \ce{MnO} (AFM)   & 3d & 7.84 & 6.98 & 4.99 & 4.49 & 8.08 & 7.07 \\
    \ce{CoO} (AFM)   & 3d & 7.12 & 6.11 & 7.46 & 5.73 & 7.02   & 6.49 \\
    \ce{NiO} (AFM)   & 3d & 14.18 & 8.01 & 14.40 & 8.89 & 7.92 & 7.24 \\
    \ce{Cu2O}        & 3d & 13.74 & 13.12 & 14.38 & 12.67  & 13.83 & 13.20 \\
    \ce{CuO} (AFM)   & 3d & 15.71 & 12.87 & 15.45 & 14.34 & 10.18 & 9.18 \\
    \midrule
    \multicolumn{8}{l}{\textbf{\textit{Alkaline earth and post-transition metal oxides}}} \\
    \midrule
    \ce{BeO}        & 2s  & 0.10 & 0.03 & 0.10 & 0.03 & 0.10 & 0.03 \\
    \ce{MgO}        & 2p & 21.39 & 20.31 & 21.63 & 20.67 & 21.39 & 20.31 \\
    \ce{CaO}        & 3p & 8.25 & 7.57 & 9.47 & 8.51 & 8.25 & 7.57 \\
    \ce{SrO}        & 4p &  5.00 & 4.42 & 6.36 & 5.01 & 5.00 & 4.42\\
    \ce{BaO}        & 5p &  4.77 & 3.92 & 4.00 & 3.70 & 4.77 & 3.92\\
    \ce{ZnO}        & 3d  & 12.34 & 11.76 & 18.72 & 18.42 & 12.33 & 11.75 \\
    \ce{CdO}        & 4d  & 11.59 & 10.99 & 12.97 & 12.56 & 11.59 & 10.98 \\
    \ce{HgO}        & 5d  & 8.41 & 7.84 & 10.31 & 9.42 & 8.30 & 7.75 \\
    \midrule
    \multicolumn{8}{l}{\textbf{\textit{Nitrides}}} \\
    \midrule
    \ce{GaN}     & 3d    & 19.78 & 19.04 & 24.56 & 23.72 & 19.78 & 19.04 \\
    \ce{AlN}     & 3p    & 0.10 & 0.05 & 0.10 & 0.05 & 0.10 & 0.05 \\
    \bottomrule
  \end{tabular}
\end{table}

\begin{table}[htbp]
  \centering
  \caption{Hubbard $U$ values for O-$2p$ and N-$2p$ shells (eV) from ACBN0@PBE with the L\"{o}wdin projector. ``l'' and ``t'' denote the ``light'' and ``tight'' numerical settings. All calculations are single-point calculations on geometries relaxed with the PBE ``tight'' settings.}
  \label{si:tab:SI_U_NM_PBE}
  \normalsize
  \setlength{\tabcolsep}{3pt}
  \renewcommand{\arraystretch}{1.05}
  \begin{tabular}{l cc cc cc}
    \toprule
    & \multicolumn{6}{c}{\textbf{ACBN0@PBE}} \\
    & \multicolumn{2}{c}{Petukhov} & \multicolumn{2}{c}{FLL} & \multicolumn{2}{c}{AMF} \\
    \cmidrule(lr){2-3} \cmidrule(lr){4-5} \cmidrule(lr){6-7}
    \textbf{Material} & \makebox[2.3em]{l} & \makebox[2.3em]{t} & \makebox[2.3em]{l} & \makebox[2.3em]{t} & \makebox[2.3em]{l} & \makebox[2.3em]{t} \\
    \midrule
    \multicolumn{7}{l}{\textbf{\textit{3d transition metal oxides}}} \\
    \midrule
    \ce{TiO2} (rut.)  & 5.40 & 4.53 & 5.57 & 4.59 & 5.40 & 4.53 \\
    \ce{TiO2} (an.)  & 5.41 & 4.54 & 5.58 & 4.61 & 5.41 & 4.54 \\
    \ce{Cr2O3} (AFM)  & 3.96 & 3.59 & 3.94 & 3.53 & 4.05 & 3.63 \\
    \ce{MnO} (AFM)    & 3.45 & 3.01 & 2.23 & 1.88 & 3.57 & 3.06 \\
    \ce{CoO} (AFM)    & 1.76 & 1.24 & 2.00 & 1.14 & 1.65   & 1.37 \\
    \ce{NiO} (AFM)    & 4.23 & 1.42 & 4.52 & 1.80 & 1.48 & 1.13 \\
    \ce{Cu2O}         & 1.99 & 1.81 & 1.46 & 0.88   & 2.15 & 1.92 \\
    \ce{CuO} (AFM)    & 4.31 & 2.67 & 4.48 & 3.33 & 2.04 & 1.52 \\
    \midrule
    \multicolumn{7}{l}{\textbf{\textit{Alkaline earth and post-transition metal oxides}}} \\
    \midrule
    \ce{BeO}          & 4.89 & 3.23 & 4.99 & 3.23 & 4.89 & 3.23 \\
    \ce{MgO}          & 6.82 & 4.57 & 7.17 & 4.63 & 6.82 & 4.57 \\
    \ce{CaO}          & 6.90 & 5.13 & 7.48 & 5.31 & 6.90 & 5.13 \\
    \ce{SrO}          & 6.16 & 5.03 & 6.66 & 5.24 & 6.16 & 5.03\\
    \ce{BaO}          & 6.64 & 5.10 & 7.29 & 5.33 & 6.64 & 5.10\\
    \ce{ZnO}          & 1.81 & 1.55 & 5.01 & 4.17 & 1.80 & 1.55 \\
    \ce{CdO}          & 3.78 & 3.22 & 5.35 & 4.45 & 3.78 & 3.21 \\
    \ce{HgO}          & 2.94 & 2.45 & 4.54 & 3.42 & 2.89 & 2.41 \\
    \ce{$\alpha$-Bi2O3} & 5.45;\,5.34;\,5.23 & 4.75;\,4.70;\,4.61 & 5.64;\,5.51;\,5.38 & 4.84;\,4.78;\,4.68 & 5.45;\,5.34;\,5.23 & 4.75;\,4.70;\,4.61 \\
    \midrule
    \multicolumn{7}{l}{\textbf{\textit{Nitrides}}} \\
    \midrule
    \ce{GaN}          & 2.32 & 1.85 & 2.53 & 2.01 & 2.32 & 1.85 \\
    \ce{AlN}          & 2.26 & 1.54 & 2.26 & 1.54 & 2.26 & 1.54 \\
    \bottomrule
  \end{tabular}
\end{table}

\begin{table}[htbp]
  \centering
  \caption{Hubbard $U$ values for the listed atomic shells (eV) from ACBN0@SCAN with the L\"{o}wdin projector. ``l'' and ``t'' denote the ``light'' and ``tight'' numerical settings. All calculations are single-point calculations on geometries relaxed with the PBE ``tight'' settings.}
  \label{si:tab:SI_U_Me_SCAN}
  \normalsize
  \setlength{\tabcolsep}{3pt}
  \renewcommand{\arraystretch}{1.05}
  \begin{tabular}{l c cc cc cc}
    \toprule
    &  & \multicolumn{6}{c}{\textbf{ACBN0@SCAN}} \\
    & & \multicolumn{2}{c}{Petukhov} & \multicolumn{2}{c}{FLL} & \multicolumn{2}{c}{AMF} \\
    \cmidrule(lr){3-4} \cmidrule(lr){5-6} \cmidrule(lr){7-8}
    \textbf{Material} & \textbf{Shell}& \makebox[2.3em]{l} & \makebox[2.3em]{t} & \makebox[2.3em]{l} & \makebox[2.3em]{t} & \makebox[2.3em]{l} & \makebox[2.3em]{t} \\
    \midrule
    \multicolumn{8}{l}{\textbf{\textit{3d transition metal oxides}}} \\
    \midrule
    \ce{TiO2} (rut.) & 3d  & 0.45 & 0.35 & 0.41 & 0.34 & 0.45 & 0.35 \\
    \ce{TiO2} (an.) & 3d & 0.47 & 0.36 & 0.42 & 0.34 & 0.47 & 0.36 \\
    \ce{Cr2O3} (AFM) & 3d & 2.21 & 2.03 & 2.13 & 1.95 & 2.30 & 2.08 \\
    \ce{MnO} (AFM)   & 3d & 7.46 & 6.62 & 5.15 & 4.43 & 7.61 & 6.68 \\
    \ce{CoO} (AFM)   & 3d & 7.94 & 5.97 & 6.74/11.05 & 5.94/5.98 & 6.67 & 6.17 \\
    \ce{NiO} (AFM)  & 3d  & 14.84 & 8.75 & 14.89 & 10.81 & 7.82 & 7.12 \\
    \ce{Cu2O}        & 3d & 13.65 & 13.05 & 14.38 & 12.67 & 13.67 & 13.08 \\
    \ce{CuO} (AFM)   & 3d & 16.09 & 13.86 & 15.91 & 14.99 & 10.55 & 9.48 \\
    \midrule
    \multicolumn{8}{l}{\textbf{\textit{Alkaline earth and post-transition metal oxides}}} \\
    \midrule
    \ce{BeO}        & 2s   & 0.09 & 0.02 & 0.09 & 0.02 & 0.09 & 0.02 \\
    \ce{MgO}        & 2p   & 21.35 & 20.29 & 21.58 & 20.63 & 21.35 & 20.29 \\
    \ce{CaO}        & 3p  & 8.00 & 7.35 & 9.39 & 8.42 & 8.00 & 7.35 \\
    \ce{SrO}        & 4p  & 5.29 & 4.65 & 5.99 & 4.64 & 5.29 & 4.65\\
    \ce{BaO}        & 5p  & 4.89 & 4.03 & 4.27 & 3.87 & 4.89 & 4.03\\
    \ce{ZnO}        & 3d  & 12.28 & 11.74 & 19.07 & 18.70 & 12.27 & 11.74 \\
    \ce{CdO}       &  4d   & 11.68 & 11.10 & 13.14 & 12.68 & 11.68 & 11.09 \\
    \ce{HgO}       &  5d    & 8.33 & 7.80 & 10.30 & 9.38 & 8.23 & 7.70 \\
    \midrule
    \multicolumn{8}{l}{\textbf{\textit{Nitrides}}} \\
    \midrule
    \ce{GaN}       & 3d  & 20.02 & 19.28 & 24.48 & 23.62 & 20.02 & 19.28 \\
    \ce{AlN}       & 3p   & 0.09 & 0.05 & 0.09 & 0.05 & 0.09 & 0.05 \\
    \bottomrule
  \end{tabular}
\end{table}

\begin{table}[htbp]
  \centering
  \caption{Hubbard $U$ values for O-$2p$ and N-$2p$ shells (eV) from ACBN0@SCAN with the L\"{o}wdin projector. ``l'' and ``t'' denote the ``light'' and ``tight'' numerical settings. All calculations are single-point calculations on geometries relaxed with the PBE ``tight'' settings.}
  \label{si:tab:SI_U_NM_SCAN}
  \normalsize
  \setlength{\tabcolsep}{3pt}
  \renewcommand{\arraystretch}{1.05}
  \begin{tabular}{l cc cc cc}
    \toprule
    & \multicolumn{6}{c}{\textbf{ACBN0@SCAN}} \\
    & \multicolumn{2}{c}{Petukhov} & \multicolumn{2}{c}{FLL} & \multicolumn{2}{c}{AMF} \\
    \cmidrule(lr){2-3} \cmidrule(lr){4-5} \cmidrule(lr){6-7}
    \textbf{Material} & \makebox[2.3em]{l} & \makebox[2.3em]{t} & \makebox[2.3em]{l} & \makebox[2.3em]{t} & \makebox[2.3em]{l} & \makebox[2.3em]{t} \\
    \midrule
    \multicolumn{7}{l}{\textbf{\textit{3d transition metal oxides}}} \\
    \midrule
    \ce{TiO2} (rut.)  & 5.54 & 4.62 & 5.72 & 4.69 & 5.54 & 4.62 \\
    \ce{TiO2} (an.)  & 5.54 & 4.64 & 5.72 & 4.71 & 5.54 & 4.64 \\
    \ce{Cr2O3} (AFM)  & 3.72 & 3.35 & 3.76 & 3.33 & 3.79 & 3.39 \\
    \ce{MnO} (AFM)    & 3.34 & 2.88 & 2.37 & 1.90 & 3.42 & 2.90 \\
    \ce{CoO} (AFM)    & 2.21 & 1.24 & 2.62 & 1.26 & 1.61 & 1.31 \\
    \ce{NiO} (AFM)    & 4.58 & 1.76 & 4.80 & 2.47 & 1.54 & 1.16 \\
    \ce{Cu2O}         & 1.97 & 1.80 & 1.46 & 0.88 & 2.09 & 1.88 \\
    \ce{CuO} (AFM)    & 4.57 & 3.02 & 4.84 & 3.63 & 2.22 & 1.63 \\
    \midrule
    \multicolumn{7}{l}{\textbf{\textit{Alkaline earth and post-transition metal oxides}}} \\
    \midrule
    \ce{BeO}          & 5.06 & 3.31 & 5.17 & 3.31 & 5.06 & 3.31 \\
    \ce{MgO}          & 7.06 & 4.70 & 7.42 & 4.76 & 7.06 & 4.70 \\
    \ce{CaO}          & 7.08 & 5.23 & 7.66 & 5.42 & 7.08 & 5.23 \\
    \ce{SrO}          & 6.33 & 5.13 & 6.81 & 5.33 & 6.33 & 5.13 \\
    \ce{BaO}          & 6.85 & 5.23 & 7.52 & 5.47 & 6.85 & 5.23 \\
    \ce{ZnO}          & 1.82 & 1.56 & 5.11 & 4.24 & 1.81 & 1.56 \\
    \ce{CdO}          & 3.91 & 3.32 & 5.50 & 4.55 & 3.91 & 3.31 \\
    \ce{HgO}          & 2.91 & 2.42 & 4.52 & 3.39 & 2.86 & 2.38 \\
    \ce{$\alpha$-Bi2O3} & 5.50;\,5.37;\,5.26 & 4.78;\,4.71;\,4.62 & 5.69;\,5.55;\,5.41 & 4.88;\,4.80;\,4.70 & 5.50;\,5.37;\,5.26 & 4.78;\,4.71;\,4.61 \\
    \midrule
    \multicolumn{7}{l}{\textbf{\textit{Nitrides}}} \\
    \midrule
\ce{GaN}          & 2.40 & 1.89 & 2.57 & 2.06 & 2.40 & 1.89 \\
    \ce{AlN}          & 2.34 & 1.58 & 2.33 & 1.57 & 2.34 & 1.58 \\
    \bottomrule
  \end{tabular}
\end{table}

\clearpage

\begingroup
\small
\setlength{\tabcolsep}{3.0pt}
\renewcommand{\arraystretch}{1.12}
\begin{longtable}{llrrrrrrrrr}
\caption{Basis dependence of ACBN0--FLL at fixed PBE/``tight''-relaxed geometries and fixed ``tight'' integration grids. L, T, and E denote the ``light'', ``tight'', and expanded orbital bases. E activates all second-tier functions on the metal and all second- and third-tier functions on O. $N_b$ is the number of basis functions retained after removal of linearly dependent combinations. $E_g$ is the band gap, $|m_M|$ is the local spin-only Mulliken moment on the metal, and $U_d$ and $U_p$ are effective Hubbard parameters for metal-$3d$ and O-$2p$. Ranges retain inequivalent site values without averaging. Occupations in the CoO rows are displayed to four decimal places, matching the Co-site values in Table~\ref{si:tab:SI_CoO_basis_sites}. $N_M^\sigma$ and $N_O^\sigma$ are occupations averaged over the orbital components of the specified shell in spin channel $\sigma$, shown for the first metal and O sites. For nonmagnetic Cu$_2$O and TiO$_2$, a single spin-summed occupation $N=2N^\uparrow$ is displayed across the two spin columns of each shell. The opposite AFM site has interchanged spin channels when equivalent. (an.) and (rut.) denote anatase and rutile. Energies are in eV and moments in $\mu_B$.}\label{si:tab:SI_basis_dependence}\\
\toprule
Material & Basis & $N_b$ & $E_g$ & $|m_M|$ & $U_d$ & $U_p$ & $N_M^\uparrow$ & $N_M^\downarrow$ & $N_O^\uparrow$ & $N_O^\downarrow$\\
\midrule
\endfirsthead
\multicolumn{11}{l}{Table~\thetable\ continued}\\
\toprule
Material & Basis & $N_b$ & $E_g$ & $|m_M|$ & $U_d$ & $U_p$ & $N_M^\uparrow$ & $N_M^\downarrow$ & $N_O^\uparrow$ & $N_O^\downarrow$\\
\midrule
\endhead
\midrule
\multicolumn{11}{r}{Continued on the next page}\\
\endfoot
\bottomrule
\endlastfoot
\multicolumn{11}{l}{\textbf{ACBN0@PBE--FLL}}\\
CoO & L & 96 & 2.596 & 2.802 & 7.54 & 2.04 & 0.9799 & 0.4385 & 0.6841 & 0.6841\\
CoO & T & 174 & 1.941 & 2.754 & 5.73 & 1.14 & 0.9183 & 0.4238 & 0.5814 & 0.5814\\
CoO & E & 241 & 1.736 & 2.825 & 5.41 & 0.94 & 0.8869 & 0.4104 & 0.5294 & 0.5294\\
Cr$_2$O$_3$ & L & 208 & 2.719 & 3.145 & 2.66 & 3.96 & 0.138 & 0.691 & 0.689 & 0.689\\
Cr$_2$O$_3$ & T & 394 & 2.508 & 3.096 & 2.44 & 3.53 & 0.128 & 0.630 & 0.635 & 0.635\\
Cr$_2$O$_3$ & E & 673 & 2.294 & 3.165 & 2.22 & 1.86 & 0.115 & 0.563 & 0.453 & 0.453\\
Cu$_2$O & L & 152 & 1.395 & 0.000 & 14.38 & 1.46 & \multicolumn{2}{c}{1.948} & \multicolumn{2}{c}{1.300}\\
Cu$_2$O & T & 238 & 1.079 & 0.000 & 12.67 & 0.88 & \multicolumn{2}{c}{1.890} & \multicolumn{2}{c}{1.190}\\
Cu$_2$O & E & 375 & 0.967 & 0.000 & 11.12 & 0.65 & \multicolumn{2}{c}{1.789} & \multicolumn{2}{c}{0.967}\\
CuO & L & 90 & 1.913 & 0.829 & 15.45 & 4.51 & 0.983 & 0.833 & 0.676 & 0.676\\
CuO & T & 158 & 1.670 & 0.825 & 14.34 & 3.33 & 0.954 & 0.812 & 0.606 & 0.606\\
CuO & E & 244 & 1.246 & 0.807 & 10.42 & 1.55 & 0.899 & 0.781 & 0.500 & 0.500\\
MnO & L & 90 & 2.014 & 4.961 & 4.99 & 2.25 & 0.963 & 0.056 & 0.711 & 0.711\\
MnO & T & 174 & 1.701 & 5.040 & 4.49 & 1.88 & 0.909 & 0.055 & 0.647 & 0.647\\
MnO & E & 282 & 1.485 & 5.314 & 4.35 & 1.14 & 0.850 & 0.047 & 0.481 & 0.481\\
NiO & L & 96 & 3.672 & 1.926 & 14.41 & 4.55 & 0.986 & 0.617 & 0.701 & 0.701\\
NiO & T & 174 & 3.075 & 1.843 & 8.89 & 1.80 & 0.934 & 0.607 & 0.593 & 0.593\\
NiO & E & 256 & 2.940 & 1.806 & 7.07 & 0.92 & 0.901 & 0.593 & 0.496 & 0.496\\
TiO$_2$ (an.) & L & 236 & 2.548 & 0.000 & 0.46 & 5.63 & \multicolumn{2}{c}{0.443} & \multicolumn{2}{c}{1.390}\\
TiO$_2$ (an.) & T & 472 & 2.350 & 0.000 & 0.37 & 4.61 & \multicolumn{2}{c}{0.395} & \multicolumn{2}{c}{1.257}\\
TiO$_2$ (an.) & E & 705 & 2.231 & 0.000 & 0.32 & 3.11 & \multicolumn{2}{c}{0.358} & \multicolumn{2}{c}{1.031}\\
TiO$_2$ (rut.) & L & 118 & 2.298 & 0.000 & 0.45 & 5.63 & \multicolumn{2}{c}{0.437} & \multicolumn{2}{c}{1.387}\\
TiO$_2$ (rut.) & T & 236 & 2.019 & 0.000 & 0.36 & 4.59 & \multicolumn{2}{c}{0.391} & \multicolumn{2}{c}{1.249}\\
TiO$_2$ (rut.) & E & 351 & 1.926 & 0.000 & 0.31 & 3.09 & \multicolumn{2}{c}{0.352} & \multicolumn{2}{c}{1.024}\\
\multicolumn{11}{l}{\textbf{ACBN0@SCAN--FLL}}\\
CoO & L & 96 & 3.377 & 2.851--2.924 & 6.73--11.08 & 2.62 & 0.9808 & 0.4315 & 0.7002 & 0.6921\\
CoO & T & 174 & 2.639 & 2.822--2.834 & 5.94--5.98 & 1.26 & 0.9209 & 0.4120 & 0.5897 & 0.5893\\
CoO & E & 242 & 2.371 & 2.854--2.855 & 5.40--5.43 & 0.98 & 0.8901 & 0.3995 & 0.5359 & 0.5361\\
Cr$_2$O$_3$ & L & 208 & 3.926 & 3.062 & 2.14 & 3.78 & 0.132 & 0.677 & 0.699 & 0.699\\
Cr$_2$O$_3$ & T & 394 & 3.758 & 2.992 & 1.95 & 3.33 & 0.122 & 0.621 & 0.643 & 0.643\\
Cr$_2$O$_3$ & E & 673 & 3.563 & 2.976 & 1.76 & 1.72 & 0.107 & 0.558 & 0.458 & 0.458\\
Cu$_2$O & L & 152 & 1.641 & 0.000 & 14.33 & 1.44 & \multicolumn{2}{c}{1.948} & \multicolumn{2}{c}{1.310}\\
Cu$_2$O & T & 238 & 1.333 & 0.000 & 12.67 & 0.88 & \multicolumn{2}{c}{1.891} & \multicolumn{2}{c}{1.198}\\
Cu$_2$O & E & 375 & 1.224 & 0.000 & 11.17 & 0.67 & \multicolumn{2}{c}{1.790} & \multicolumn{2}{c}{0.973}\\
CuO & L & 90 & 2.193 & 0.934 & 15.91 & 4.83 & 0.983 & 0.814 & 0.688 & 0.688\\
CuO & T & 158 & 1.943 & 0.931 & 14.99 & 3.63 & 0.955 & 0.795 & 0.615 & 0.615\\
CuO & E & 244 & 1.766 & 0.985 & 12.19 & 2.04 & 0.902 & 0.757 & 0.508 & 0.508\\
MnO & L & 90 & 2.706 & 5.004 & 5.18 & 2.40 & 0.964 & 0.047 & 0.721 & 0.721\\
MnO & T & 174 & 2.376 & 5.016 & 4.43 & 1.90 & 0.912 & 0.048 & 0.655 & 0.655\\
MnO & E & 282 & 2.157 & 5.228 & 4.21 & 1.12 & 0.855 & 0.040 & 0.486 & 0.486\\
NiO & L & 96 & 4.359 & 1.965 & 14.97 & 4.84 & 0.987 & 0.613 & 0.711 & 0.711\\
NiO & T & 174 & 3.760 & 1.951 & 10.81 & 2.47 & 0.937 & 0.593 & 0.602 & 0.602\\
NiO & E & 256 & 3.604 & 1.872 & 8.19 & 1.26 & 0.904 & 0.578 & 0.503 & 0.503\\
TiO$_2$ (an.) & L & 236 & 2.976 & 0.000 & 0.43 & 5.76 & \multicolumn{2}{c}{0.425} & \multicolumn{2}{c}{1.406}\\
TiO$_2$ (an.) & T & 472 & 2.768 & 0.000 & 0.34 & 4.71 & \multicolumn{2}{c}{0.379} & \multicolumn{2}{c}{1.269}\\
TiO$_2$ (an.) & E & 705 & 2.628 & 0.000 & 0.30 & 3.16 & \multicolumn{2}{c}{0.345} & \multicolumn{2}{c}{1.039}\\
TiO$_2$ (rut.) & L & 118 & 2.707 & 0.000 & 0.41 & 5.77 & \multicolumn{2}{c}{0.419} & \multicolumn{2}{c}{1.404}\\
TiO$_2$ (rut.) & T & 236 & 2.390 & 0.000 & 0.34 & 4.69 & \multicolumn{2}{c}{0.375} & \multicolumn{2}{c}{1.262}\\
TiO$_2$ (rut.) & E & 351 & 2.277 & 0.000 & 0.29 & 3.15 & \multicolumn{2}{c}{0.339} & \multicolumn{2}{c}{1.033}\\
\end{longtable}
\begin{table}[htbp]
\centering
\caption{Site-resolved Co parameters for the completed CoO basis tests. The two Co sites are reported separately because the SCAN solutions do not retain exact equivalence of their projected occupations and interactions. The basis labels and units are as in Table~\ref{si:tab:SI_basis_dependence}.}\label{si:tab:SI_CoO_basis_sites}
\begin{tabular}{llrrrrr}
\toprule
Parent & Basis & Co site & $U_{\mathrm{Co}-3d}$ & $N_{\mathrm{Co}-3d}^\uparrow$ & $N_{\mathrm{Co}-3d}^\downarrow$ & $m_{\mathrm{Co}}$\\
\midrule
PBE & L & 1 & 7.54 & 0.9799 & 0.4385 & 2.802\\
PBE & L & 2 & 7.53 & 0.4386 & 0.9799 & -2.802\\
PBE & T & 1 & 5.73 & 0.9183 & 0.4238 & 2.754\\
PBE & T & 2 & 5.73 & 0.4238 & 0.9183 & -2.754\\
PBE & E & 1 & 5.41 & 0.8869 & 0.4104 & 2.825\\
PBE & E & 2 & 5.41 & 0.4104 & 0.8869 & -2.825\\
SCAN & L & 1 & 6.73 & 0.9808 & 0.4315 & 2.851\\
SCAN & L & 2 & 11.08 & 0.4215 & 0.9828 & -2.924\\
SCAN & T & 1 & 5.94 & 0.9209 & 0.4120 & 2.822\\
SCAN & T & 2 & 5.98 & 0.4108 & 0.9206 & -2.834\\
SCAN & E & 1 & 5.40 & 0.8901 & 0.3995 & 2.855\\
SCAN & E & 2 & 5.43 & 0.3997 & 0.8901 & -2.854\\
\bottomrule
\end{tabular}
\end{table}
\endgroup
\clearpage

\begin{table}[htbp]
\centering
\caption{Selected bond lengths (in \AA) for the clean monolayer slab model of the $\beta$-NiOOH(001) surface and adsorbate-covered structures calculated with different exchange-correlation functionals. The first three distances describe the local Ni--$O_{\mathrm{s}}$ coordination around the reactive surface oxygen site. The remaining columns list adsorbate-related distances, where $O_{\mathrm{s}}$ is the surface oxygen site, and $O_1$, $O_2$ denote the proximal and distal oxygen atoms of the OOH intermediate. The large PBE value of $d(O_{\mathrm{s}}-O_1)$ for the $\mathrm{*OOH}$ intermediate reflects displacement of the OOH fragment from the surface oxygen site and formation of an H--$O_{\mathrm{s}}$ bond, as described in the main text. HSE06, PBE0, and rSCAN stabilize an intact O-O-O motif.}
\label{si:tab:oer_intermediate_bond_lengths}
\footnotesize
\setlength{\tabcolsep}{3pt}
\renewcommand{\arraystretch}{1.08}
\begin{tabular}{llccccccccc}
\toprule
State & Functional & \multicolumn{3}{c}{$d(\mathrm{Ni}-O_{\mathrm{s}})$} & $d(O_{\mathrm{s}}-H)$ & $d(O_{\mathrm{s}}-O_1)$ & $d(O_1-H)$ & $d(O_1-O_2)$ & $d(O_2-H)$ & $d(H-O_{\mathrm{s}})$ \\
\cmidrule(lr){3-5}
 & & 1 & 2 & 3 & & & & & & \\
\midrule
clean & HSE06 & 1.90 & 2.03 & 2.14 & - & - & - & - & - & - \\
clean & PBE0  & 1.90 & 2.03 & 2.14 & - & - & - & - & - & - \\
clean & rSCAN & 1.93 & 2.00 & 2.12 & - & - & - & - & - & - \\
clean & PBE   & 1.94 & 2.03 & 2.14 & - & - & - & - & - & - \\
\midrule
$\mathrm{*H}$ & HSE06 & 1.92 & 1.98 & 2.11 & 0.96 & - & - & - & - & - \\
$\mathrm{*H}$ & PBE0  & 1.92 & 1.98 & 2.11 & 0.96 & - & - & - & - & - \\
$\mathrm{*H}$ & rSCAN & 2.00 & 2.00 & 2.02 & 0.96 & - & - & - & - & - \\
$\mathrm{*H}$ & PBE   & 2.01 & 2.01 & 2.02 & 0.97 & - & - & - & - & - \\
\midrule
$\mathrm{*O}$ & HSE06 & 1.93 & 2.08 & 2.25 & - & 1.29 & - & - & - & - \\
$\mathrm{*O}$ & PBE0  & 1.94 & 2.10 & 2.28 & - & 1.28 & - & - & - & - \\
$\mathrm{*O}$ & rSCAN & 1.96 & 2.01 & 2.03 & - & 1.32 & - & - & - & - \\
$\mathrm{*O}$ & PBE   & 1.95 & 2.02 & 2.04 & - & 1.34 & - & - & - & - \\
\midrule
$\mathrm{*OH}$ & HSE06 & 1.89 & 2.08 & 2.08 & - & 1.42 & 0.98 & - & - & - \\
$\mathrm{*OH}$ & PBE0  & 1.89 & 2.08 & 2.08 & - & 1.42 & 0.98 & - & - & - \\
$\mathrm{*OH}$ & rSCAN & 1.91 & 1.97 & 2.17 & - & 1.45 & 1.00 & - & - & - \\
$\mathrm{*OH}$ & PBE   & 1.93 & 1.98 & 2.16 & - & 1.48 & 1.01 & - & - & - \\
\midrule
$\mathrm{*OOH}$ & HSE06 & 1.93 & 2.05 & 2.10 & - & 1.41 & - & 1.39 & 0.98 & - \\
$\mathrm{*OOH}$ & PBE0  & 1.93 & 2.05 & 2.11 & - & 1.41 & - & 1.39 & 0.98 & - \\
$\mathrm{*OOH}$ & rSCAN & 1.94 & 2.02 & 2.06 & - & 1.45 & - & 1.41 & 0.99 & - \\
$\mathrm{*OOH}$ & PBE   & 1.90 & 1.95 & 1.96 & - & 3.09 & - & 1.32 & 1.11 & 1.36 \\
\bottomrule
\end{tabular}
\end{table}

\begin{table}[htbp]
  \centering
  \caption{Ranges of converged effective Hubbard parameters $U=\bar U-\bar J$ for the clean $\beta$-NiOOH(001) slab, oxygen-evolution intermediates, and oxygen-containing molecular references calculated with ACBN0@rSCAN using the L\"{o}wdin projector and the Petukhov mixing parameter for the double-counting term (eV). The O-$2p$ values are reported separately for the eight oxygen atoms of the NiOOH slab and for the oxygen atoms of the adsorbate or molecular reference.}
  \label{si:tab:SI_surface_U}
  \normalsize
  \setlength{\tabcolsep}{5pt}
  \renewcommand{\arraystretch}{1.08}
  \begin{tabular}{l ccc ccc}
    \toprule
    & \multicolumn{3}{c}{\textbf{``light''}} & \multicolumn{3}{c}{\textbf{``tight''}} \\
    \cmidrule(lr){2-4} \cmidrule(lr){5-7}
    \textbf{System} & $U_{\mathrm{Ni}\text{-}3d}$ & $U_{\mathrm{O}\text{-}2p}^{\mathrm{slab}}$ & $U_{\mathrm{O}\text{-}2p}^{\mathrm{ads./ref.}}$ & $U_{\mathrm{Ni}\text{-}3d}$ & $U_{\mathrm{O}\text{-}2p}^{\mathrm{slab}}$ & $U_{\mathrm{O}\text{-}2p}^{\mathrm{ads./ref.}}$ \\
    \midrule
    Clean slab       & 5.49--6.29 & 2.91--3.24 & --         & 4.38--4.67 & 2.02--2.08 & --         \\
    $\mathrm{*OH}$   & 4.38--4.52 & 2.67--3.12 & 2.91       & 3.84--4.13 & 1.86--2.39 & 2.21       \\
    $\mathrm{*O}$    & 3.92--5.35 & 2.74--3.10 & 3.24       & 4.03--4.43 & 1.93--2.65 & 2.65       \\
    $\mathrm{*OOH}$  & 3.72--4.21 & 2.79--3.26 & 3.04,3.23 & 3.55--3.77 & 1.88--2.47 & 2.27,2.55 \\
    $\mathrm{H_2O}$  & --         & --         & 6.69       & --         & --         & 4.49       \\
    $\mathrm{O_2}$   & --         & --         & 7.66       & --         & --         & 6.13       \\
    \bottomrule
  \end{tabular}
\end{table}

\begin{table}[!t]
  \centering
\caption{Decomposition of the OER reaction-step deviations from HSE06/``tight'' for rSCAN and ACBN0@rSCAN into slab ($\delta E_{i}^{\mathrm{slab}}$) and molecular-reference ($\delta E_{i}^{\mathrm{mol}}$) contributions. Here $E(\mathrm{*})$ is the total energy of the clean slab, and $E(\mathrm{*X})$ is the total energy of the slab with adsorbed $X$ ($X=\mathrm{O},\mathrm{OH},\mathrm{OOH}$), not the energy of an isolated adsorbate. $E(\mathrm{H}_2)$, $E(\mathrm{H_2O})$, and $E(\mathrm{O}_2)$ refer to isolated gas-phase molecules. For steps 1--4, the slab terms are $E(\mathrm{*OH})-E(\mathrm{*})$, $E(\mathrm{*O})-E(\mathrm{*OH})$, $E(\mathrm{*OOH})-E(\mathrm{*O})$, and $E(\mathrm{*})-E(\mathrm{*OOH})$. The corresponding molecular terms are $\tfrac{1}{2}E(\mathrm{H}_2)-E(\mathrm{H_2O})$, $\tfrac{1}{2}E(\mathrm{H}_2)$, $\tfrac{1}{2}E(\mathrm{H}_2)-E(\mathrm{H_2O})$, and $\tfrac{1}{2}E(\mathrm{H}_2)+E(\mathrm{O}_2)$, respectively. Each tabulated contribution is the difference between the corresponding term calculated with the indicated method and with HSE06/``tight''. Their sum is the net reaction-step deviation, $\delta\Delta E_i^{\mathrm{el}}$. The contributions and net deviations are evaluated from unrounded energies and rounded separately to 0.001~eV for display, so the displayed components need not sum exactly to the displayed net deviation. The ``light''-basis/``tight''-grid calculations retain the ``light'' basis functions while using the ``tight'' integration grids.}
  \label{si:tab:SI_oer_decomposition}
  \small
  \setlength{\tabcolsep}{7pt}
  \renewcommand{\arraystretch}{1.00}
  \begin{tabular}{llrrrr}
    \toprule
    \textbf{Method} & \textbf{Settings} & \textbf{Step} & $\delta E_i^{\mathrm{slab}}$ & $\delta E_i^{\mathrm{mol}}$ & $\delta\Delta E_i^{\mathrm{el}}$ \\
    \midrule
    rSCAN & ``tight'' & 1 & -1.420 &  1.641 &  0.221 \\
          &       & 2 & -0.142 & -0.035 & -0.177 \\
          &       & 3 & -1.400 &  1.641 &  0.241 \\
          &       & 4 &  2.962 & -3.324 & -0.362 \\
    \midrule
    ACBN0@rSCAN & ``light'' & 1 & -1.688 &  1.510 & -0.179 \\
                &       & 2 &  0.300 & -0.027 &  0.273 \\
                &       & 3 & -1.785 &  1.510 & -0.275 \\
                &       & 4 &  3.173 & -3.001 &  0.172 \\
    ACBN0@rSCAN & ``light'' basis/``tight'' grids & 1 & -1.695 &  1.511 & -0.184 \\
                &       & 2 &  0.323 & -0.025 &  0.297 \\
                &       & 3 & -1.819 &  1.511 & -0.308 \\
                &       & 4 &  3.191 & -3.006 &  0.185 \\
    ACBN0@rSCAN & ``tight'' & 1 & -1.489 &  1.640 &  0.151 \\
                &       & 2 &  0.018 & -0.035 & -0.017 \\
                &       & 3 & -1.610 &  1.640 &  0.030 \\
                &       & 4 &  3.080 & -3.312 & -0.232 \\
    \bottomrule
  \end{tabular}
\end{table}

\clearpage

\end{document}